# Graph-based Automated Discovery of Concise Soil Hydraulic Functions from Data: Beyond Empirical Formulations

**Hao Xu[1,2], Jinshen Sun[3,4], Yuntian Chen[1,5,*] and Dongxiao Zhang[1,*]**

[1] Zhejiang Key Laboratory of Industrial Intelligence and Digital Twin, Eastern Institute of Technology, Ningbo, Zhejiang 315200, China.

[2] Department of Electrical Engineering, Tsinghua University, Beijing 100084, P. R. China.

[3] CNPC Engineering Technology R&D Company Limited, Beijing 102206, P. R. China.

[4] School of Petroleum Engineering, China University of Petroleum (East China), Qingdao 266580, P. R. China.

5 Ningbo Institute of Digital Twin, Eastern Institute of Technology.

*Corresponding author: Y. Chen (ychen@eitech.edu.cn); D. Zhang (dzhang@eitech.edu.cn)

**Key Points:**

- Graph-based equation discovery identified a new soil hydraulic model from the data used to develop the Mualem-van Genuchten model.
- A compact tangent-based form is discovered to analytically describe soil water retention and hydraulic conductivity functions.
- Across soils in the UNSODA dataset, the model achieved lower mean conductivity error than the Mualem-van Genuchten model and its variants.

## Abstract

Soil hydraulic functions underpin water flow and transport modeling in the vadose zone, yet they are commonly represented by expert-designed formulations such as the Mualem–van Genuchten (MvG) model. Because these formulations rest on predefined functional assumptions, balancing predictive accuracy and parametric compactness across soil textures remains challenging. Here we introduce a graph-based equation discovery framework that distills soil hydraulic functions from experimental data. Reapplied to the original datasets that motivated the MvG parameterization, the framework recovers a concise water retention function together with an analytical unsaturated hydraulic conductivity expression whose mathematical structure departs from classical empirical forms. Evaluated across soils of contrasting texture, the discovered model predicts unsaturated hydraulic conductivity more accurately than MvG and its common variants, while its fitted parameters retain interpretable correlations with measurable soil physical properties. These findings suggest that data-driven discovery can yield compact, analytically tractable constitutive relations that move beyond fixed empirical formulations.

## Plain Language Summary

Soil holds water in its pores and releases it as it dries. How fast water can still move through partly dry soil governs irrigation, groundwater recharge, and flood generation. This property—unsaturated hydraulic conductivity—is difficult and expensive to measure directly, so it is usually estimated from water retention data, which are easier to collect. The formulas used for that conversion, such as the widely used Mualem–van Genuchten (MvG) model, were built through expert intuition and manual curve fitting. We developed an automated equation discovery method that systematically searches a vast space of possible formulas, trained on the same laboratory measurements that originally motivated the MvG model. It identified a compact formula built on the tangent function, a mathematical form largely overlooked in conventional soil hydraulic descriptions. When tested on soils with different textures, the new model predicted unsaturated hydraulic conductivity with lower average error than the standard model and several of its variants. It remained comparably simple and, importantly, reduced the frequency of very large prediction errors. This shows that data-driven discovery can yield compact, analytically tractable formulas that improve on fixed empirical formulations.

## 1 Introduction

Soil hydraulic functions, most notably the soil-water retention curve (SWRC) and the hydraulic conductivity function (HCF), are indispensable for describing water flow in variably saturated porous media and therefore underpin a wide range of applications in hydrology, agriculture, and geoscience (Patil & Singh, 2016). Given the challenges of first-principles derivation in heterogeneous, multiscale systems, classical soil hydraulic models remain largely phenomenological, combining physical intuition with empirical fitting to experimental observations (Assouline & Or, 2013; Fredlund & Xing, 1994). For example, the widely-used Mualem–van Genuchten (MvG) model was historically formulated through a heuristic process of observation and iterative trial-and-error (van Genuchten, 1980). While this Galilean-style heuristic modeling has historically been effective, the process is inherently inefficient and expertise-intensive, relying on subjective intuition and scientific serendipity. More generally, empirical modeling struggles to strike an optimal balance between structural parsimony and predictive accuracy, posing a major challenge for soil water modeling.

With advancements in machine learning, automatic discovery of equations directly from data has gained increasing attention, which usually relies on evolutionary computation and symbolic regression (Bongard & Lipson, 2007; Schmidt & Lipson, 2009) . For dynamical systems, sparse regression identifies governing terms from predefined candidate libraries, with extensions to partial differential equations (Brunton et al., 2016; Rudy et al., 2017). Genetic programming provides a flexible alternative by evolving expression structures to identify nonlinear system models, including soil-moisture dynamics (Im et al., 2021). Related stochastic frameworks jointly optimize equation structures and coefficients for PDE identification (Im et al., 2023a), while neural-guided genetic programming uses learned expression distributions to initialize candidate populations (Mundhenk et al., 2021). These developments have supported equation discovery in spatial heterogeneous problems (Im et al., 2023b; Xu et al., 2025).

However, soil hydraulic function discovery requires a shared constitutive form that accommodates differences among soils through soil-specific parameters. It also involves a mathematical transformation between retention and conductivity. In capillary models such as those of Mualem and Burdine, conductivity prediction depends on inversion and integration of the retention relation (Burdine, 1953; Mualem, 1976), so candidate selection must consider these operations alongside fitting accuracy. Therefore, conventional model discovery methods with predefined libraries (Chang & Zhang, 2019; Song et al., 2022) are difficult to apply directly to this problem, as soil hydraulic functions are often too intricate to be captured through manual term enumeration (Song et al., 2024). Consequently, a central question arises: Can data-driven equation discovery identify closed-form constitutive models for the SWRC, thereby enabling more accurate HCF predictions that generalize across diverse conditions?

Here, we apply a graph-based model discovery framework for the identification of soil hydraulic functions (Figure 1a), which is different from conventional tree-based symbolic regression methods (Chen et al., 2022; Cranmer, 2023). It utilizes graphs to represent equations, where nodes correspond to variables and operators and edges capture computational dependencies. It can therefore represent soil-specific parameters in the equation structure through edge features. On this basis, a tangent-based function is identified directly from the same experimental data with the classical MvG model. We further evaluate the discovered functions on soil samples from the Unsaturated Soil Hydraulic Database (UNSODA) (Nemes et al., 2001), and the discovered model achieves more accurate predictions of unsaturated hydraulic conductivity compared with variants of the MvG model with fewer free parameters. This demonstrates that data-driven model discovery can uncover explicit constitutive relations beyond established empirical forms.

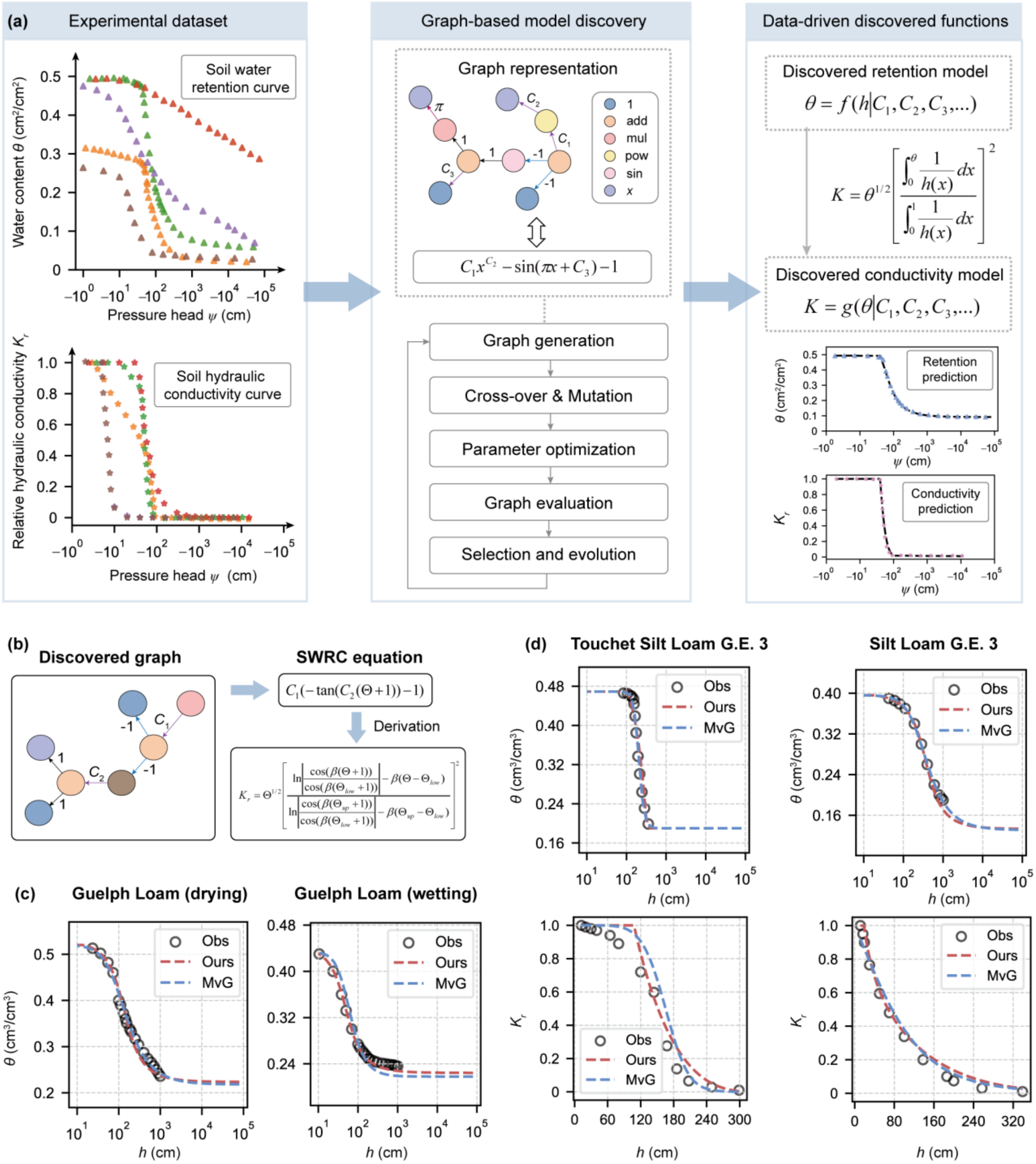

**Figure 1. Graph-based discovery of soil hydraulic functions.** (a) Workflow from experimental retention and conductivity data to graph-based equation discovery and model prediction. (b) Discovered graph structure, corresponding soil water retention equation, and analytically derived conductivity function. (c) Retention-curve predictions for Guelph Loam under drying and wetting conditions. (d) Retention and conductivity predictions for two soils, compared with observations and the Mualem-van Genuchten (MvG) model.

## 2 Data and Methods

### 2.1 Datasets

The equation discovery is based on a set of classic soil water retention curves and hydraulic conductivity measurements originally compiled and used in the development of the Mualem–van Genuchten (MvG) model (van Genuchten, 1980; Mualem, 1976). This dataset includes six curves that collectively capture a wide range of textural and hydrological characteristics: Beit Netofa Clay, Hygiene Sandstone, Silt Loam G.E. 3, Touchet Silt Loam G.E. 3, and Guelph Loam under drying and wetting conditions. Each sample is associated with a set of data points describing the soil water retention curve. Except for Guelph loam, all other cases include unsaturated hydraulic conductivity curves. Cross-soil applicability was evaluated using soil samples from the UNSODA. More details about these datasets are provided in Text S1. For each soil, model parameters were calibrated using only the retention measurements. The resulting function was then used to predict relative hydraulic conductivity.

### 2.2 Graph-based equation discovery framework

For the data-driven equation discovery, a graph-based representation of equations is adopted to express formulas with soil-specific parameters. In the graph, nodes correspond to either a variable or an operator and directed edges define the computational dependencies between them (Xu et al., 2026) . Addition and multiplication nodes are allowed to connect to multiple edges. Edge features are introduced to encode fixed constants and the soil-specific parameters, which allows a single equation structure to be evaluated jointly across multiple soils while retaining parameter values specific to each soil.

The discovery process is implemented through genetic programming. Initially, several candidate graphs are randomly generated. Structural constraints are introduced to regulate the equation form and prevent the construction of syntactically complex expressions. Then, crossover and mutation are applied to explore new graphs. Crossover swaps randomly selected subgraphs between two graphs. Mutation comprises three operations: edge feature mutation (replacing an edge feature), subgraph mutation (replacing a subgraph), and subgraph deletion (replacing a subgraph with a variable or constant node). Each generated graph is then evaluated using a loss function:

$$Loss = \frac{1}{M}\sum_{i=1}^{M}\left(\frac{1}{N_i}\sum_{j=1}^{N_i}(f(x_j^i \mid C_1^i, C_2^i, ..., C_K^i) - y_j^i)^2\right) + \varepsilon \cdot N_C \tag{1}$$

Here, $f$ denotes the candidate equation, $M$ is the number of cases, $N_i$ is the number of observation records for case $i$, and $N_C$ represents the number of undetermined soil-specific parameters $C^i$ . $(x_j^i, y_j^i)$ are the values of the independent and dependent variables in case $i$, and $\varepsilon$ is the complexity penalty. Here, a low loss value indicates good performance. For each case, the soil-specific parameters are estimated using the L-BFGS optimization algorithm (Liu & Nocedal, 1989). On this basis, the top 50% of the graphs in the current population are retained as parents for the next generation, whereas the remaining individuals are replaced by newly generated candidates.

To maintain diversity and avoid local minima, a periodic extinction strategy is applied, whereby only the five best-performing individuals are preserved and all others are replaced. The highest-performing graphs in the final generation are returned as the discovered equations. The detailed settings are provided in Text S2.

## 3 Results

### 3.1. Data-driven discovery of soil-water retention curve functions

On the basis of data from six curves in Mualem's original dataset, we first apply the graph-based equation discovery framework to identify a new, concise SWRC function that captures the relationship between volumetric water content ($\theta$) and pressure head ($\psi$). We define $h=|\psi|=-\psi>0$ as the magnitude of the pressure head under unsaturated conditions. Rather than searching directly for $\theta=f(h)$, we search for an inverse relation $1/h=g(\theta)$. This formulation follows the Mualem capillary model (Mualem, 1976), in which the calculation of relative hydraulic conductivity ($K_r$) involves integration with respect to $1/h$, and therefore favors retention functions that yield explicit conductivity expressions. We use the dimensionless pressure head $\boldsymbol{\hat{h} = h/h_0}$, with a reference head of $h_0$=1 cm.

Effective saturation Θ is utilized as the input variable, which is defined as:

$$\Theta = \frac{\theta - \theta_r}{\theta_s - \theta_r} \tag{2}$$

where $\theta_s$ and $\theta_r$ indicate saturated and residual water contents, respectively. For numerical stability, the discovery target was set to $100/\boldsymbol{\hat{h}}$. For the graph-based representation, the operator set to generate potential functions is (*add*, *mul*, *exp*, *pow*, *log*, *sin*, *cos*, *tan*), with no more than two soil-specific parameters. The optimization trajectory shows clear loss convergence. Model selection was based on the joint consideration of fitting accuracy and integrability. Although the rank-1 function attains the lowest fitness, its form is not explicitly integrable. We therefore adopted the rank-2 form, which is both concise and integrable (Figure S1). The optimal function is written as:

$$\frac{1}{\hat{h}} \times 100 = C_1(-\tan(C_2(\Theta+1)) - 1) \tag{3}$$

where $C_1$ and $C_2$ are soil-specific parameters to be determined (Figure 1b). Tangent-containing candidates appeared in four of the five independent runs (Table S1). Let $A=C_1/100$, $\beta=C_2$, Eq. (3) can be written as:

$$\frac{1}{\hat{h}} = A(-\tan(\beta(\Theta+1)) - 1) \tag{4}$$

Here, A and $\beta$ are dimensionless parameters. From Eq. (4), the effective saturation Θ can be easily solved as a function of $h$:

$$\Theta = \frac{1}{\beta}\arctan\left(-\left(\frac{1}{A\hat{h}}+1\right)\right)-1 \tag{5}$$

The derivation of the inverse function is given in Text S3. Although the inverse-tangent relation is multi-valued, the solution is restricted to its principal branch, which means that the selected tangent interval is ($-\pi/2$, $-\pi/4$). Therefore, the admissible parameter ranges are $A>0$ and $\beta<0$. Under this condition, the function remains monotonic over the fitted range of pressure heads and is free of singularities. The tangent-family structure identified here differs from the functional forms of the classical unimodal soil water retention models. Existing models have largely relied on a limited set of functional families, most notably power-law forms such as the Brooks-Corey (Brooks & Corey, 1964) and Campbell models (Campbell, 1974), sigmoidal forms such as the MvG model (van Genuchten, 1980), and lognormal-transform-based expressions such as the Kosugi model (Kosugi, 1994). Despite their differences, these models all reflect a relatively narrow range of human-designed hypotheses. By contrast, automated model discovery can explore a wider functional space, thereby uncovering constitutive forms that are difficult to anticipate through expert intuition alone. As illustrated in Figure 1c, the discovered model precisely characterizes the soil-water retention curves, encompassing both drying and wetting cycles.

### 3.2. Derivation of hydraulic conductivity function based on the discovered model

Following identification of the soil water retention function, we derived the corresponding relative hydraulic conductivity $K_r$ using the generalized Mualem capillary formulation:

$$K_r = \Theta^{1/2}\left[\int_{\Theta_{low}}^{\Theta}\frac{1}{\hat{h}(x)}dx \Big/ \int_{\Theta_{low}}^{\Theta_{up}}\frac{1}{\hat{h}(x)}dx\right]^2 \tag{6}$$

This formulation is intended primarily for soils in which conductivity is governed by capillary pore-size effects. To avoid singular behavior at the theoretical endpoints, this formation is the approximation of relative hydraulic conductivity with finite integration limits, $\Theta_{low}$ and $\Theta_{up}$, corresponding to pressure heads of $\hat{\boldsymbol{h}}=10^5$ and $\hat{\boldsymbol{h}}=1$, respectively. Detailed analysis of the integration bounds is provided in Text S4. The standard Mualem capillary model can be seen as the special case with $\Theta_{low}=0$ and $\Theta_{up}=1$. Then, Eq. (4) can be analytically integrated as:

$$\int\frac{1}{\hat{h}(x)}dx = \frac{A}{\beta}\ln\left|\cos(\beta(x+1))\right| - Ax + C \tag{7}$$

Substituting Eq. (7) into Eq. (6) and simplifying yields:

$$K_r = \Theta^{1/2}\left[\frac{\ln\left|\frac{\cos(\beta(\Theta+1))}{\cos(\beta(\Theta_{low}+1))}\right| - \beta(\Theta-\Theta_{low})}{\ln\left|\frac{\cos(\beta(\Theta_{up}+1))}{\cos(\beta(\Theta_{low}+1))}\right| - \beta(\Theta_{up}-\Theta_{low})}\right]^2 \tag{8}$$

Once $A$ and $\beta$ are estimated from the soil water retention data, the soil hydraulic conductivity function $K_r$ can be calculated directly without numerical integration from Eq. (8). As shown in Figure 1d, the derived hydraulic conductivity function predicts $K_r$ accurately over a wide range of pressure heads.

Notably, obtaining an explicit hydraulic conductivity function is intrinsically difficult, because conductivity depends on pore structure, connectivity and tortuosity, which are only partially encoded in water retention behavior. Simple SWRC forms can preserve analytical tractability, but often with limited predictive fidelity. More flexible empirical modifications (Durner, 1994; Fredlund & Xing, 1994) can improve fitting accuracy, yet frequently lead to conductivity functions that require numerical evaluation. The tractability is important in the implementation in large-scale hydrological modeling or analytical derivations (Ippisch et al., 2006). For example, solutions of the Richards equation need tractable first-order derivatives and parameter sensitivities. Historically, the Gardner-Russo model (Gardner, 1958; Russo, 1988) has been preferred for analytical solutions due to its simplicity, despite its limited predictive accuracy. While the MvG model addresses this fitting limitation, the derivation of the derivatives and sensitivities required for robust numerical and stochastic modeling can become cumbersome under certain parameterizations (Zhang, 2002). In the discovered model, the mapping between $h$ and $\Theta$ is fully explicit and analytically invertible. Moreover, the parameters $A$ and $\beta$ are structurally decoupled from the exponent system, avoiding the nested exponent terms in MvG model. As demonstrated in Text S5, exact analytical derivatives and sensitivities for both $\theta$ and $K_r$ yield compact expressions, facilitating analytical, numerical, and stochastic modeling without sacrificing accuracy. The analysis of the pore-size distribution implied by the discovered function is provided in Text S6.

### 3.3. Extension of the discovered model and evaluation across UNSODA soils

To assess the cross-soil applicability of the discovered model, we evaluated it using 249 samples from UNSODA. For each soil, the model parameters, including the residual water content $\theta_r$, were calibrated using only the water retention data, and the corresponding relative hydraulic conductivity was predicted from the closed-form expression. $\theta_s$ is fixed at the measured saturated or near-saturated water content. All retention observations, including those outside the range represented by the untruncated relation, were retained in the objective function. The implementation applies a [0,1] truncation only to the effective saturation predicted by the model. During the parameter optimization, constraints were imposed on $\beta$, and the same principal branch was used for all soil samples. Here, the UNSODA dataset is divided into a development set (20%), a validation set (20%) and a test set (60%). On the development set, the discovered model achieved a mean RMSE of 0.026 for $\theta$ and 3.128 for $\ln(K_r)$. However, for some soils, the optimized $\theta_r$ fell outside physically plausible ranges and the conductivity prediction errors remained large. These limitations reflect the restricted flexibility of the functional form in representing the full water-retention curve, as detailed in Text S3 and Figure S2.

Motivated by these observations, we introduced a minimal extension of the discovered tangent-based relation:

$$\frac{1}{\hat{h}} = A(-\tan^{p}(\beta(\Theta+1))-1), \quad p \in \{1,3\} \tag{9}$$

When $p$=1, this expression reduces exactly to the originally discovered model Eq. (4). Allowing $p$ to take values of 1 or 3 adds a single discrete shape choice (Figure S3), while retaining the tangent structure and the main continuous fitting parameters $A$ and $\beta$. These odd powers preserve branch consistency during inversion and remain analytically integrable within the Mualem framework. The inverse retention function of Eq. (9) on the physically admissible branch can be written explicitly as:

$$\Theta(\hat{h}) = \frac{1}{\beta}\arctan\left[\left(-\left(\frac{1}{A\hat{h}}+1\right)\right)^{1/p}\right]-1 \tag{10}$$

Defining

$$G(\Theta) = \int A[-\tan^{p}(\beta(\Theta+1))-1]d\Theta, \tag{11}$$

For $p$∈{1,3}, Eq. (11) can be written in a unified analytical form:

$$G(\Theta) = -\frac{A(p-1)}{4\beta}\tan^{2}(\beta(\Theta+1)) - \frac{A(p-2)}{\beta}\ln\left|\cos(\beta(\Theta+1))\right| - A\Theta, \tag{12}$$

This equation unifies the separately evaluated integrals for $p$=1 and $p$=3, but does not apply to general values of $p$. The relative hydraulic conductivity can be given by:

$$K_r(\Theta) = \Theta^{1/2}\left[\frac{G(\Theta)-G(\Theta_{low})}{G(\Theta_{up})-G(\Theta_{low})}\right]^{2} \tag{13}$$

When $p$=1, Eq. (13) reduces to Eq. (8). Therefore, the modified model preserves analytical simplicity and closed-form conductivity prediction while providing greater flexibility.

Figure 2a compares the original and modified models in fitting the soil water retention curve on the validation dataset. Even though the original model already achieved high accuracy, the modified model further reduced the RMSE of $\theta$ by 15.1% overall. The modified model reduced the RMSE of ln($K_r$) by 11.8% (Figure 2b). Representative examples further show that the extended model alters the dry-end extrapolation behavior and achieves more accurate conductivity predictions in cases where the original relation lacks sufficient flexibility (Figure S4). Importantly, direct model discovery on UNSODA failed to identify a candidate that improved retention and relative hydraulic conductivity predictions simultaneously over Eq. (9) (Table S2). These results highlight the importance of data quality and informativeness for model discovery and the influence of the discovery objective (Text S7).

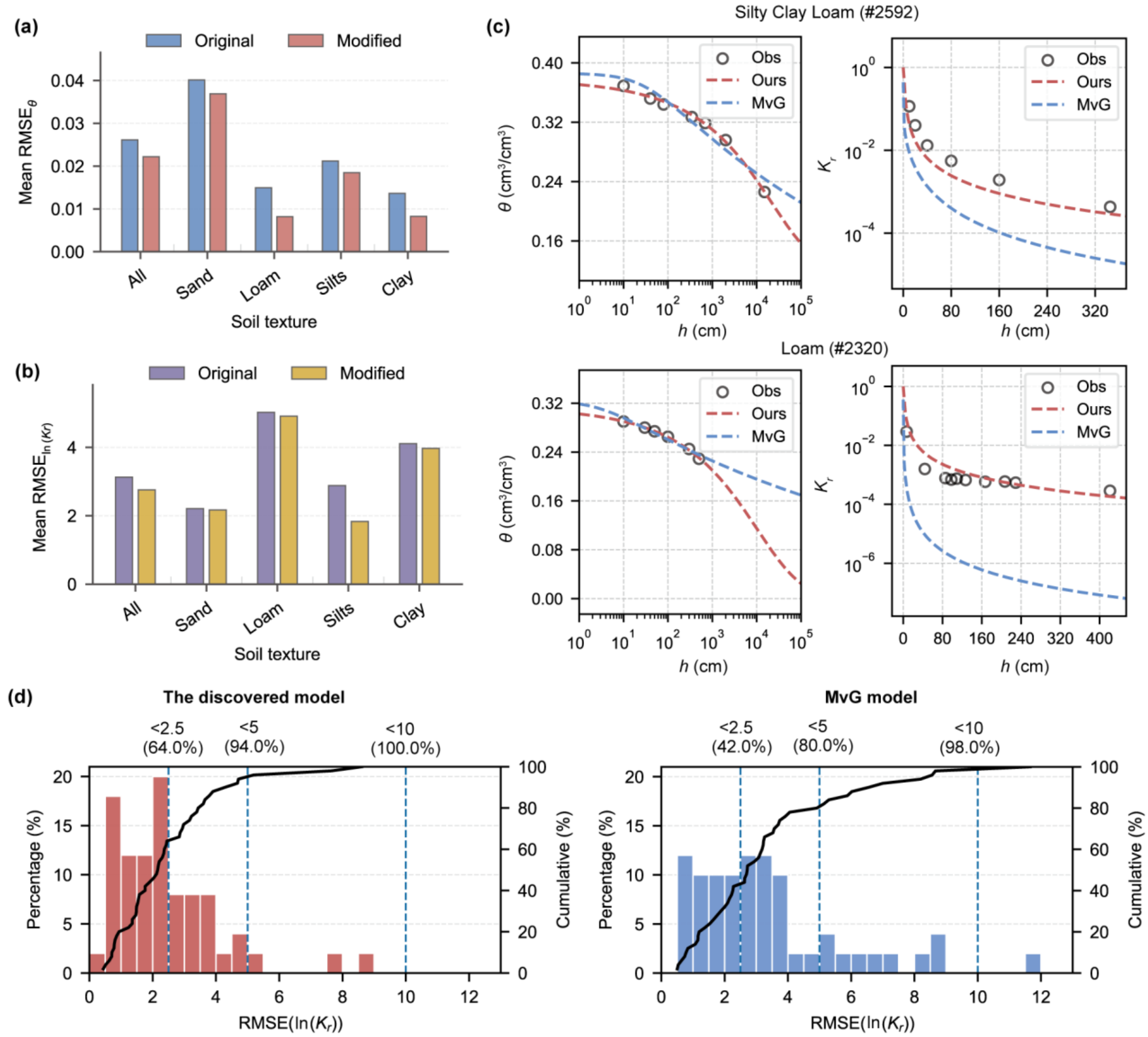

**Figure 2. Performance of the discovered soil hydraulic model across UNSODA soils in the validation set.** (a, b) Mean retention and conductivity errors of the original and modified discovered models across soil texture classes. (c) Representative retention and conductivity predictions for two soils. (d) Distributions and cumulative percentages of conductivity prediction errors for the discovered and Mualem-van Genuchten (MvG) models.

## 4 Discussion

### 4.1. Comparison with classical Mualem–van Genuchten model

To better evaluate the performance of the modified model, we first compare it with the widely used classical Mualem–van Genuchten (MvG) model (van Genuchten, 1980) using the validation set from UNSODA. The retention function in the MvG model is written as:

$$\Theta = \left[ \frac{1}{1+(\alpha h)^n} \right]^m \tag{14}$$

$$\frac{1}{h} = \alpha[\Theta^{-1/m} - 1]^{-1/n}, \quad \text{with } m = 1 - 1/n \tag{15}$$

$$K_r(\Theta) = \Theta^{1/2}[1-(1-\Theta^{1/m})^m]^2 \tag{16}$$

From the perspective of structural simplicity, the MvG model involves three parameters: $\alpha$, $n$, and $m$, with the relation $m$=1−1/$n$. The modified model also contains three parameters, but with $p$ restricted to a discrete choice between 1 and 3. This distinction reflects a difference between equation structures guided by human intuition and those uncovered through data-driven model discovery. Both models generally reproduced the observed retention curves, but the conductivity predictions of the tangent-based model are better. Representative cases indicate that this improvement arises mainly from conductivity behavior near saturation (Figure 2c). It also reduced the frequency of large prediction errors (Figure 2d). In the MvG model, if the optimized parameter $n<2$ and particularly $1.0<n<1.3$, the hydraulic conductivity function may exhibit a sharp decline at pressure heads slightly below saturation (Luckner et al., 1989; Vogel et al., 2000), which can lead to significant prediction errors (Vogel et al., 2000). Such parameter values occurred frequently in the evaluated soils and contributed to several large conductivity errors (Figure S5). By contrast, the modified model produced smoother near-saturated behavior and more stable conductivity estimates. Although it showed slightly higher retention errors (Figure S6), it provided more accurate and robust conductivity predictions across diverse soil textures.

4.2 Comparison with existing soil water models

We further compared the modified model with representative variants derived from the MvG framework, including the air-entry modified MvG model (Schaap & van Genuchten, 2006), the Durner bimodal MvG model (Durner, 1994), and the PDI-MvG model (Peters et al., 2024). Detailed settings and descriptions of these models are provided in Text S8. These models extend the classical MvG model to address near-saturated behavior, multimodal pore-size distributions, and hydraulic conductivity prediction. The benchmark results on the test set from UNOSDA are summarized in Table 1. The air-entry modified MvG model introduces an additional air-entry suction parameter to represent the near-saturated plateau, reducing the error in ln($K_r$) by 9.4% (from 3.003 to 2.720). Even relative to this improved baseline, the proposed method further lowers the error by 10.7% (from 2.720 to 2.430). The Durner bimodal MvG model enables a better description of heterogeneous pore structures, but at the cost of increased model complexity. The recent PDI-MvG model provides improved performance for conductivity prediction, although its predictive formulation is more complicated. As shown in Table S3, the modified model uses fewer fitted parameters and involves only one discrete parameter $p$. Paired comparisons and the Wilcoxon signed-rank test indicate that the modified model achieves a relative improvement in the challenging task of predicting $K_r$ compared with these MvG variants (Figure S7). The mean and median absolute errors, together with the interquartile ranges, further support this conclusion (Figure S8). Our model uses finite integration bounds, $\Theta_{\text{low}}$ and $\Theta_{\text{up}}$ in Eq. (13), with the corresponding head cutoffs treated as fixed hyperparameters. For a fair comparison, Table

S4 evaluates classical MvG and PDI-MvG using the same head cutoffs. Truncating the Mualem integral modifies near-saturation conductivity and improves the predictive accuracy of these models. Nevertheless, the proposed model retains the lowest overall RMSE, supporting the benefit of the discovered functional form beyond integral truncation alone.

**Table 1 Comparisons of predictive performance for hydraulic conductivity between different models.** $RMSE_{ln(Kr)}$ denotes the root mean squared error for logarithmic relative hydraulic conductivity. Reported percentages indicate the fraction of soil samples meeting the specified error thresholds.

| **Models** | Classical MvG | Air-entry modified MvG | Durner bimodal MvG | PDI-MvG | Ours |
|---|---|---|---|---|---|
| **$RMSE_{ln(Kr)}$** | 3.003 | 2.720 | 3.271 | 2.660 | **2.430** |
| **$RMSE_{ln(Kr)}$<2.5** | 61.1% | 61.7% | 54.4% | 52.3% | **65.8%** |
| **$RMSE_{ln(Kr)}$<5** | 84.6% | 89.3% | 82.6% | 92.6% | **92.6%** |
| **$RMSE_{ln(Kr)}$<10** | 96.0% | 98.0% | 96.0% | 100% | **100%** |

4.3 Correlation analysis of the fitted parameters

The fitted parameters exhibited systematic relationships with soil texture and hydraulic properties. Coarse-textured soils, particularly sands, more frequently selected $p$=1, whereas fine-textured soils, such as loams and clays, favored $p$=3 (Figure S9). The continuous parameters also varied across the soil texture space. Both $\log_{10}(A)$ and $\beta$ were generally higher in sand-rich soils and lower in silt-dominated soils, and their distributions differed significantly among texture classes (Figure 3a and 3b). For $\beta$, the Kruskal–Wallis test yielded $H$=38.49, p-value=$2.21\times10^{-8}$, indicating statistically significant differences in the distributions among groups. Similarly, for $\log_{10}(A)$, the Kruskal–Wallis test ($H$=99.54, p-value=$1.95\times10^{-21}$) supported the existence of intergroup variations. Interestingly, both $\log_{10}(A)$ and $\beta$ decline as the soil particle size diminishes from sandy to clayey textures (Figure 3c). In contrast, the optimized residual water content $\theta_r$ increases with finer textural classes (Figure 3d). From the pairwise normalized Wasserstein distances matrix (Figure 3e), it is discovered that soil types with similar textural compositions tend to exhibit comparable parameter distributions. Pairwise comparisons of soil hydraulic parameters across texture groups (Table S5) further show that $A$ differed mainly between Sand and the other groups. $\beta$ was higher in Sand than in Silts and Clay. All pairwise differences in $\theta_r$ were significant. Among the fitted parameters, $\log_{10}(A)$ showed the clearest relationships with independent hydraulic descriptors (Figure 3f and Figure S10). This pattern suggests that $A$ mainly reflects an effective pore-size or air-entry scale controlling drainage intensity, while parameter $p$ primarily governs the overall shape regime of the retention curve.

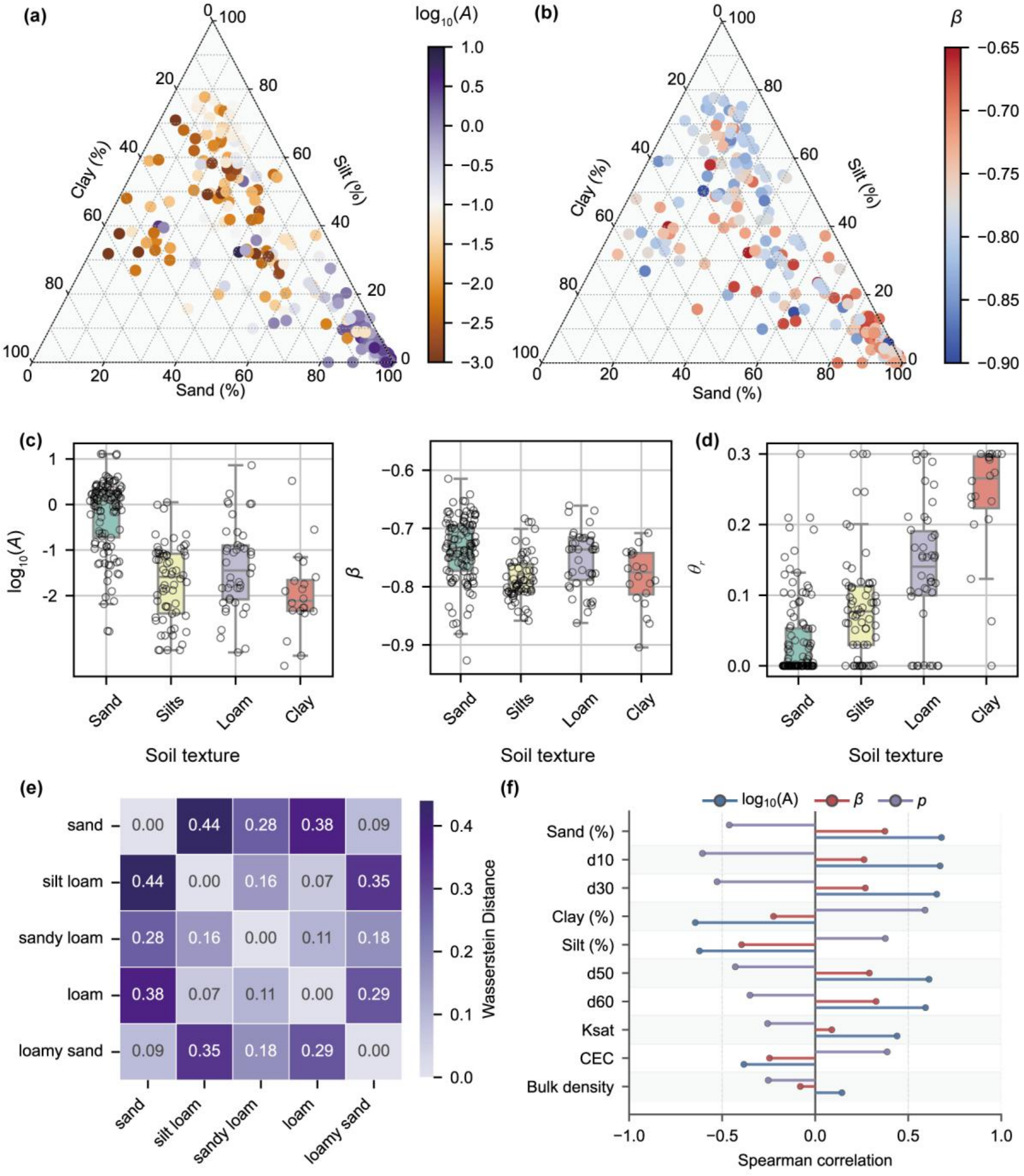

**Figure 3. Relationships between fitted model parameters and soil properties.** (a, b) Distributions of $\log_{10}(A)$ and $\beta$ across the soil texture triangle. (c, d) Parameter distributions for major soil texture classes. (e) Pairwise Wasserstein distances between $\log_{10}(A)$ distributions. (f) Spearman correlations between fitted model parameters and soil physical descriptors.

## 5 Conclusions

In this study, we used a graph-based equation discovery framework to identify concise soil hydraulic functions directly from experimental data. The discovered relation is explicitly invertible and analytically integrable within the Mualem capillary framework, allowing the corresponding relative hydraulic conductivity function to be expressed in closed form. An extension guided by mathematical considerations improved the flexibility of the original equation while preserving its compact analytical structure. Evaluation on UNSODA database further confirms the predictive ability of the model, which accurately predicts unsaturated hydraulic conductivity behaviors across a wide range of soil textures and conditions. Statistical analysis revealed that the soil-specific parameters in the discovered model are associated with soil textures and properties.

More broadly, the results demonstrate that equation discovery can complement conventional model development methods. Existing machine-learning approaches usually learn mappings between soil properties and hydraulic response (Araya & Ghezzehei, 2019; Gupta et al., 2021; Mozaffari et al., 2025). Although such models demonstrate high predictive accuracy, they cannot provide explicit functions, which poses a practical barrier to their deployment in resource-constrained environments. In contrast, the proposed framework can identify constitutive forms that may be overlooked by predefined model families. Its ability to learn a shared equation structure with soil-specific parameters also supports the use of heterogeneous observations from multiple soils. The resulting symbolic expressions can be incorporated directly into analytical, numerical, and stochastic hydrological models.

Despite these advantages, several limitations remain. Requiring analytical integrability restricts the admissible search space and may exclude functions needed to represent irregular hydraulic behavior. The tangent formulation also requires appropriate parameter and branch constraints to avoid singularities. Finally, because conductivity is derived through a Mualem-type capillary model, applicability to structured, dual-porosity, or strongly heterogeneous soils requires further evaluation. Despite these limitations, the study provides a compact alternative for soil hydraulic modeling and illustrates the potential of interpretable data-driven discovery for developing geoscientific constitutive relations.

**Acknowledgments**

This work is supported and partially funded by the National Natural Science Foundation of China (Grant No. 52288101, 12501744, 12572266), Postdoctoral Fellowship Program and China Postdoctoral Science Foundation (Grant No. BX20250063, 2024M761535), the National Key Research and Development Program (2024YFF1500600), and the Yongjiang Talent Program of Ningbo (2022A-242-G). This work is supported by the High Performance Computing Centers at Eastern Institute of Technology, Ningbo, and Ningbo Institute of Digital Twin. We thank Prof. Li Ren and Prof. Liangsheng Shi for their constructive suggestions.

**Conflict of Interest Disclosure**

The authors declare no conflicts of interest relevant to this study.

## Availability Statement

The data and source code used to produce the results and figures presented in this study are publicly available in Zenodo (Xu et al., 2026) at https://doi.org/10.5281/zenodo.21815847 under the Apache license. The code is also openly developed and maintained at https://github.com/woshixuhao/Discovery_of_unsaturated_soil_model.

Supporting Information for

# Graph-based automated discovery of concise soil hydraulic functions from data: Beyond Empirical Formulations

Hao Xu[1,2], Jinshen Sun[3,4], Yuntian Chen[1,5,*], and Dongxiao Zhang[1,*]

[1] Zhejiang Key Laboratory of Industrial Intelligence and Digital Twin, Eastern Institute of Technology, Ningbo, Zhejiang 315200, China

[2] Department of Electrical Engineering, Tsinghua University, Beijing 100084, P. R. China.

[3] CNPC Engineering Technology R&D Company Limited, Beijing 102206, P. R. China

[4] School of Petroleum Engineering, China University of Petroleum (East China), Qingdao 266580, P. R. China

[5] Ningbo Institute of Digital Twin, Eastern Institute of Technology, Ningbo, Zhejiang 315200, P. R. China

*****Corresponding author: Y. Chen (ychen@eitech.edu.cn); D. Zhang (dzhang@eitech.edu.cn)

**Contents of this file**

**Introduction**

This Supporting Information contains supplementary methodological descriptions, figures, and tables that support the analyses presented in the main text.

**Text S1**

The UNSODA evaluation cohort comprised 249 soils selected from 790 records represented in the available laboratory-data exports. Eligibility initially required at least five water-retention observations and five hydraulic-conductivity observations, leaving 336 soils. Saturated water content, $\theta_s$, was taken from the database when available. Otherwise, the water content at the lowest measured suction head was used, provided that this head was below $5\,\mathrm{cm}$; that observation was subsequently removed from the retention fitting data. This requirement retained 258 soils, of which 249 had at least six remaining retention observations. Selection depended on data availability and observation counts, independently of model fitting performance. The selected cohort contained 2,767 retention observations and 4,238 conductivity observations. Retention records contained 6–52 observations per soil, with a median of 10, whereas conductivity records contained 5–89 observations, with a median of 11. Retention fitting used 2,738 finite, paired observations with positive suction heads. Conductivity was normalized as $K_r = K/K_s$. Missing entries were removed, while the remaining observations retained their original sampling density.

Texture summaries used database-provided labels aggregated into broad analysis groups comprising Sand (*n*=110), Loam (*n*=42), Silts (*n*=62), Clay (*n*=20), and Unknown (*n*=15). These groups represent the classifications used for subgroup analysis. Errors were calculated separately for each soil and then averaged with equal weight across soils. Specifically, the conductivity error for soil $i$ was:

$$\mathrm{RMSE}_{\ln K_r} = \left[\frac{1}{N_K}\sum_{j=1}^{N_K}\left\{\ln\left[\max\left(K_{r,j}^{\mathrm{pred}},\varepsilon\right)\right] - \ln\left[\max\left(K_{r,j}^{\mathrm{obs}},\varepsilon\right)\right]\right\}^2\right]^{1/2}. \tag{S.1}$$

where the $\varepsilon = 10^{-8}$ . The reported mean was

$$\overline{\mathrm{RMSE}}_{\ln K_r} = \frac{1}{N}\sum_{i=1}^{N}\mathrm{RMSE}_{\ln K_r,i}, \tag{S.2}$$

where *N* is the number of soils in the evaluated subset. Thus, each soil contributed equally despite differences in observation counts. For analyses using the current partition, the cohort comprised 50 development soils, 50 validation soils, and 149 test soils, with records sharing a location, site identifier, or publication identifier assigned to the same partition.

**Text S2**

Equation discovery employed genetic programming with a population of 300 graphs evolved over 150 generations. The best-performing half of the population formed the parent pool, and each offspring underwent crossover. Mutation probabilities were defined per invocation: edge-feature mutation was attempted with probability 0.5 per edge, while subgraph modification and deletion had probabilities of 0.5 and 0.2, respectively. Node-level mutation had probability 0.3 per node. Additional mutation was performed during heuristic structural control, which targeted fewer than 30 edges and seven free symbols, including the input variable; graph depth remained unconstrained. When the best fitness remained exactly unchanged for five consecutive generations, the best individual was retained and the remaining population was regenerated. Soil-specific coefficients were estimated separately for each candidate using unconstrained L-BFGS. The overall multiplicative coefficient was initialized at unity, and other active coefficients were sampled from {-2,-1,1,2}. Twenty initializations were evaluated per candidate and soil, followed by a final optimization from the initialization yielding the lowest objective value. Fitness was minimized as the mean of the soil-specific mean squared errors.

Operator safeguards combined limited symbolic preprocessing with fitness penalization. The adjustable coefficients allowed operator arguments to change during fitting and potentially enter their valid domains at the observed inputs. Candidates yielding NaN fitness received penalties of $10^8$ during initialization and offspring evaluation. Subsequent evaluation on the UNSODA soils used bounded L-BFGS-B optimization with 84 initializations for each of $p$=1 and $p$=3. The bounds were $A \in [10^{-8}, 10^{6}]$, $\beta \in [-1,0)$, and $\theta_r \in [10^{-8}, \theta_{\min} - 10^{-8})$, where $\theta_{\min}$ denotes the minimum observed water content. Retention predictions used the arctangent inverse with branch correction and a real cube root for $p$=3. Effective saturation was clipped to $[10^{-8}, 1 - 10^{-8}]$. Conductivity predictions were clipped to [0,1], and a floor of $10^{-8}$ was applied for logarithmic error calculations.

**Text S3**

In the main text, the discovered soil-water retention function is written as:

$$\frac{1}{\hat{h}} = A(-\tan(\beta(\Theta+1))-1) \tag{S.3}$$

In this section, the derivation of its inverse function is examined in detail, together with the mathematical conditions that accompany the inversion. When $A$>0 and $\hat{h}$>0, this function can be converted into:

$$\tan(\beta(\Theta+1)) = -1 - \frac{1}{A\hat{h}} \tag{S.4}$$

Define

$$v(h;A) = -1 - \frac{1}{A\hat{h}}, \tag{S.5}$$

when $A$>0 and $\hat{h}$>0, $v$<-1. Therefore, the complete set of periodic solutions of the tangent inversion is given by:

$$\beta(\Theta+1) = \arctan v + k\pi, \quad k \in \mathrm{Z}. \tag{S.6}$$

In the main text, the branch corresponding to $k$=0 is selected, from which Eq. (5) in the main text is obtained. Given $v$<-1, the associated angular interval is:

$$-\frac{\pi}{2} < \beta(\Theta+1) < -\frac{\pi}{4} \tag{S.7}$$

In this interval, the inverse function can be calculated:

$$\Theta = \frac{1}{\beta}\arctan\left(-\left(\frac{1}{A\hat{h}}+1\right)\right)-1 \tag{S.8}$$

According to the definition of $\Theta$, $\Theta$>0, so $\beta$ is required to be negative. Therefore, the constraints of the parameters $A$ and $\beta$ are $A$>0 and $\beta$<0.

From the inverse function, the mathematical limit can be calculated as:

$$\lim_{\hat{h}\to\infty} \Theta(\hat{h}) = -\frac{\pi}{4\beta} - 1 = \Theta_{\text{dry}}, \qquad \lim_{\hat{h}\to 0^+} \Theta(\hat{h}) = -\frac{\pi}{2\beta} - 1 = \Theta_{\text{wet}}. \tag{S.9}$$

If all $\hat{h}$>0 is required to correspond to physical saturation, it is necessary that $\Theta_{\text{dry}} \geq 0$, $\Theta_{\text{wet}} \leq 1$. When $\beta$<0, it is required that $\beta \geq -\pi/4$ and $\beta \leq -\pi/4$, which means that $\beta = -\pi/4$ is the unique value that satisfies the complete-range condition. Under this condition, $\Theta_{\mathrm{dry}}=0$, $\Theta_{\mathrm{wet}}=1$. All finite positive heads correspond to $0<\Theta<1$, with the two endpoints attained only as limiting values. Let the range of heads in actual use be:

$$0<H_{\min}\le\hat{h}\le H_{\max}<\infty. \tag{S.10}$$

Define

$$q(\hat{h};A)=\arctan\left(-1-\frac{1}{A\hat{h}}\right) \tag{S.11}$$

Then,

$$\Theta(\hat{h})=\frac{q(\hat{h};A)}{\beta}-1. \tag{S.12}$$

Because $q$ increases with $\hat{h}$ while $\beta<0$, $\Theta$ decreases with increasing $h$. It is therefore sufficient to enforce the endpoint conditions $\Theta(H_{\max})\geq 0$, $\Theta(H_{\min})\leq 1$ to guarantee $0\leq\Theta(h)\leq 1$ throughout the interval. Rearrangement then yields:

$$q(H_{\max};A)\leq\beta\leq\frac{q(H_{\min};A)}{2}. \tag{S.13}$$

Monotonicity of $\Theta$ can be derived by the differentiation:

$$\frac{d\Theta}{d\hat{h}}=\frac{A}{\beta\left[1+2A\hat{h}+2(A\hat{h})^2\right]}<0 \tag{S.14}$$

Therefore, when $A>0$, $\beta<0$, $\theta_s>\theta_r$, $\frac{d\theta}{d\hat{h}}=(\theta_s-\theta_r)\frac{d\Theta}{d\hat{h}}<0$. Thus, the monotonicity of the water-retention curve on the selected branch is mathematically guaranteed. Since $q$ belongs to $(-\pi/2, -\pi/4)$, $\cos(q)>0$. Therefore, $\ln|cos(q)|$ remain well defined. Within the prescribed branch and over the range of finite positive heads, the inverted water-retention function is free of interior tangent poles.

In this function, for $-\pi/2<\beta<-\pi/4$, physical saturation holds on $0<\hat{h}\leqslant\hat{h}_{\mathrm{dry},0}$ and drier conditions may yield negative saturation. For $\beta=-\pi/4$, physical saturation holds for all $\hat{h}>0$. For $-\pi/4<\beta<-\pi/8$, physical saturation holds on $\hat{h}\geqslant\hat{h}_{\mathrm{wet},1}$ and wetter conditions may yield saturation above unity. For the remaining negative values of $\beta$, no finite positive head corresponds to $0\leqslant\Theta\leqslant 1$. Here, $\hat{h}_{\mathrm{dry},0}=\frac{1}{A[-\tan^p\beta-1]}$, $\hat{h}_{\mathrm{wet},1}=\frac{1}{A[-\tan^p(2\beta)-1]}$. Distribution of $\Theta_{\mathrm{wet}}$ and $\Theta_{\mathrm{dry}}$ across 249 soil samples in UNSODA is presented in Figure. S2. After regression, the curve obtained from the equation may fail to cover the full physically admissible range [0,1], and may yield theoretical values $\Theta_{\mathrm{wet}}>1$ or $\Theta_{\mathrm{dry}}<0$. In practical use, the predicted results are therefore truncated to the interval [0,1].

For $p=3$, the derivation is similar. For

$$\frac{1}{\hat{h}}=A\left[-\tan^3(\beta(\Theta+1))-1\right], \tag{S.15}$$

the inverse function is:

$$\Theta(\hat{h})=\frac{1}{\beta}\arctan\left[-\left(1+\frac{1}{A\hat{h}}\right)^{1/3}\right]-1. \tag{S.16}$$

Since the real cubic function is strictly increasing, $\tan^3 u<-1$ is equivalent to $\tan u<-1$. Thus, the derivation of the effective branch and of the dry and wet endpoints proceeds as in the case $p=1$. Define

$$\omega=-\left(1+\frac{1}{A\hat{h}}\right)^{1/3}, \tag{S.17}$$

the derivative is:

$$\frac{d\Theta}{d\hat{h}}=\frac{1}{3\beta A\hat{h}^2\omega^2(1+\omega^2)}<0 \tag{S.18}$$

Thus, the function guarantees monotonicity by construction. The constraints that enforce a finite admissible range remain identical to those for $p$=1; only the optimized parameter values differ.

**Text S4**

Our model computes $K_r$ using truncated integration in Eq. (13) in the main text, with $\Theta_{\text{low}}$ and $\Theta_{\text{up}}$ corresponding to $\hat{h}$=$10^5$ and 1, respectively. At the upper bound, $K_r = \sqrt{\Theta_{\text{up}}}$, so the finite-bound expression is used as an approximation to $K/K_s$. Across the UNSODA soils, $\sqrt{\Theta_{\text{up}}}$>0.99 for 64.26% of samples and exceeds 0.9 for 94.78%. UNSODA conductivity observations are normalized by the reported saturated conductivity $K_s$ and compared directly with model predictions. Finite-head observations outside the integration interval are retained for error evaluation. Model saturations are clipped to (0,1) and predicted $K_r$ values to [0,1]. Sensitivity of relative hydraulic conductivity predictions to the integration bounds is summarized in Table S6. The wet and dry head cutoffs are denoted by $H_w$ and $H_d$, respectively. It is discovered that increasing $H_d$ changes the mean RMSE by less than 0.1%. In contrast, varying $H_w$ produces modest changes. Among the investigated combinations, $H_w$=1 yields the lowest mean RMSE.

**Text S5**

The analytical derivation of constitutive derivatives and parameter sensitivities is a fundamental requirement for robust hydrological modeling across scales, since the analytical and numerical solution of Richards equation relies on well-behaved, tractable derivatives. At the same time, their explicit availability offers a stringent test of the discovered soil hydraulic functions: a model that fits the data well but does not admit tractable sensitivity expressions is of limited use for subsequent theoretical development. The following analysis therefore evaluates the discovered constitutive relations in terms of their suitability for downstream analytical, numerical or stochastic modeling [37]. We start from the pressure-head form of the Richards equation,

$$\frac{\partial\theta}{\partial t}=\nabla\cdot[K(\nabla\psi+e_z)], \tag{S.19}$$

where $\theta$ is the volumetric water content, $\psi$ is the pressure head, $K$ is the unsaturated hydraulic conductivity, and $\mathbf{e}_z$ denotes the unit vector in the vertical direction. For the proposed constitutive relation, the normalized saturation $\Theta$ is defined through

$$\theta=\theta_r+(\theta_s-\theta_r)\Theta, \tag{S.20}$$

and the retention relation is postulated as

$$\frac{1}{\hat{h}}=A[-\tan^p(\beta(\Theta+1))-1],\ p\in\{1,3\} \tag{S.21}$$

Here, for computational and notational convenience in the model, we define $h=|\psi|=-\psi>0$ as the positive magnitude of the pressure head under unsaturated conditions. The corresponding relative conductivity is obtained through the Mualem-type construction

$$K_r(\Theta)=\Theta^{1/2}\left[\frac{\int_{\Theta_{low}}^{\Theta}\frac{1}{\hat{h}(\chi)}\,d\chi}{\int_{\Theta_{low}}^{\Theta_{up}}\frac{1}{\hat{h}(\chi)}\,d\chi}\right]^2, \tag{S.22}$$

where $\Theta_{\mathrm{low}}$ and $\Theta_{\mathrm{up}}$ are fixed lower and upper cutoff values introduced to regularize the integration bounds in practical applications. For stochastic modeling, it is convenient to decompose the conductivity as $K=K_sK_r$, or equivalently, $\ln K=Y_s+\ln K_r$, $Y_s:=\ln K_s$. The main technical task is therefore to derive the following quantities in closed form:

$$\frac{\partial\theta}{\partial\hat{h}},\ \frac{\partial\theta}{\partial A},\ \frac{\partial\theta}{\partial\beta},\ \frac{\partial lnK_r}{\partial\hat{h}},\ \frac{\partial lnK_r}{\partial A},\ \frac{\partial lnK_r}{\partial\beta}. \tag{S.23}$$

The logarithmic sensitivity expressions derived below apply to the interior of the admissible domain, where A>0, $\beta<0, \Theta>0$, and $\Theta_{\mathrm{low}}<\Theta\le\Theta_{\mathrm{up}}$. All tangent arguments must remain within the selected branch $\left(-\frac{\pi}{2},-\frac{\pi}{4}\right)$with $N(\Theta)\neq 0$ and $D\neq 0$. These conditions ensure that $K_r>0$ and its logarithmic derivatives are well defined. At the lower integration bound, $N(\Theta_{\mathrm{low}})=0$ by construction, giving $K_r(\Theta_{\mathrm{low}})=0$. Consequently, $\ln K_r$ and its sensitivities are undefined at this endpoint, and corresponding equations apply only above the lower bound. This singularity arises from taking the logarithm of a vanishing conductivity. The conductivity itself remains continuous and approaches zero at the cutoff. Numerical evaluation therefore uses $K_r=0$ directly at the lower bound and evaluates logarithmic sensitivities only within the admissible interior. The same restriction applies to the corresponding expressions for $p$=3. The parameter sensitivities presented here are derived with the saturation integration bounds held fixed.

**(1) First-order constitutive sensitivities for $\theta$ and $K_r$ when $p$=1**

Here, we introduce the auxiliary variable

$$v(\hat{h};A) = -\frac{1}{A\hat{h}}-1. \tag{S.24}$$

Then the retention relation becomes

$$\tan\big(\beta(\Theta+1)\big)=v. \tag{S.25}$$

Under the single-branch assumption,

$$\beta(\Theta+1)=\arctan v, \tag{S.26}$$

and hence the normalized saturation can be written explicitly as:

$$\Theta\big(\hat{h};A,\beta\big)=\frac{1}{\beta}\arctan\left(-\frac{1}{A\hat{h}}-1\right)-1. \tag{S.27}$$

This explicit inversion is the first major advantage of the tan-based model, since the map $h\mapsto\Theta$ does not require numerical inversion.

We now derive the water-capacity coefficient. We have

$$\frac{\partial\theta}{\partial\hat{h}}=(\theta_s-\theta_r)\frac{\partial\Theta}{\partial\hat{h}}. \tag{S.28}$$

Therefore, it is sufficient to compute $\partial\Theta/\partial h$ combined with Eq. (S.27):

$$\frac{\partial\Theta}{\partial\hat{h}}=\frac{1}{\beta}\cdot\frac{1}{1+v^2}\cdot\frac{\partial v}{\partial\hat{h}}. \tag{S.29}$$

From Eq. (S.24), its derivative with respect to $h$ is

$$\frac{\partial \nu}{\partial \hat{h}}=\frac{1}{A\hat{h}^2}. \tag{S.30}$$

Substituting this result into the previous expression yields

$$\frac{\partial \Theta}{\partial \hat{h}}=\frac{1}{\beta A\hat{h}^2(1+\nu^2)}. \tag{S.31}$$

Hence the moisture-capacity coefficient is

$$C_{\hat{h}}:=\frac{\partial \theta}{\partial \hat{h}}=(\theta_s-\theta_r)\frac{1}{\beta A\hat{h}^2(1+\nu^2)}. \tag{S.32}$$

If $A$ is treated as a random parameter, we similarly obtain

$$\frac{\partial \Theta}{\partial A}=\frac{1}{\beta}\frac{1}{1+\nu^2}\frac{\partial \nu}{\partial A}. \tag{S.33}$$

From Eq. (S.24), its derivative with respect to $A$ is

$$\frac{\partial \nu}{\partial A}=\frac{1}{A^2\hat{h}}. \tag{S.34}$$

Therefore

$$C_A:=\frac{\partial \theta}{\partial A}=(\theta_s-\theta_r)\frac{\partial \Theta}{\partial A}=(\theta_s-\theta_r)\frac{1}{\beta A^2\hat{h}(1+\nu^2)}. \tag{S.35}$$

Finally, because $\nu$ does not depend on $\beta$,

$$\frac{\partial \Theta}{\partial \beta}=\frac{\partial}{\partial \beta}\left(\frac{1}{\beta}arctan\nu-1\right)=-\frac{1}{\beta^2}arctan\nu=-\frac{\Theta+1}{\beta}. \tag{S.36}$$

Consequently,

$$C_\beta:=\frac{\partial \theta}{\partial \beta}=-(\theta_s-\theta_r)\frac{\Theta+1}{\beta}. \tag{S.37}$$

The three capacity coefficients required for stochastic linearization are therefore

$$\begin{aligned} C_h&=(\theta_s-\theta_r)\frac{1}{\beta A\hat{h}^2(1+\nu^2)},\\ C_A&=(\theta_s-\theta_r)\frac{1}{\beta A^2\hat{h}(1+\nu^2)},\\ C_\beta&=-(\theta_s-\theta_r)\frac{\Theta+1}{\beta}. \end{aligned} \tag{S.38}$$

In the discovered soil hydraulic model, the $K_r$ is represented as:

$$K_r=\Theta^{1/2}\left[\frac{\ln\left|\frac{\cos(\beta(\Theta+1))}{\cos(\beta(\Theta_{low}+1))}\right|-\beta(\Theta-\Theta_{low})}{\ln\left|\frac{\cos(\beta(\Theta_{up}+1))}{\cos(\beta(\Theta_{low}+1))}\right|-\beta(\Theta_{up}-\Theta_{low})}\right]^2 \tag{S.39}$$

For notational compactness, define

$$\begin{aligned} N(\Theta)&:=ln\left|\frac{\cos(\beta(\Theta+1))}{\cos(\beta(\Theta_{low}+1))}\right|-\beta(\Theta-\Theta_{low})\\ D(\Theta)&:=ln\left|\frac{\cos(\beta(\Theta_{up}+1))}{\cos(\beta(\Theta_{low}+1))}\right|-\beta(\Theta_{up}-\Theta_{low}) \end{aligned} \tag{S.40}$$

Then the relative conductivity becomes

$$K_r(\Theta)=\Theta^{1/2}\left[\frac{N(\Theta)}{D}\right]^2. \tag{S.41}$$

Taking logarithms gives the especially convenient expression:

$$\ln K_r = \frac{1}{2} ln\Theta + 2ln|N| - 2ln|D|. \tag{S.42}$$

Because $D$ is independent of h and $A$ when $\Theta_{low}$, $\Theta_{up}$, and $\beta$ are fixed, the derivatives of $\ln K_r$ are substantially simpler than the derivatives of $K_r$ itself.

We first differentiate $N(\Theta)$ with respect to $\Theta$:

$$\frac{\partial N}{\partial \Theta} = \frac{\partial}{\partial \Theta} ln|cos(\beta(\Theta+1))| - \beta = -\beta tan\big(\beta(\Theta+1)\big) - \beta. \tag{S.43}$$

Combined with Eqs. (S.24) and (S.25):

$$\frac{\partial N}{\partial \Theta} = \frac{\beta}{A\hat{h}}. \tag{S.44}$$

Combining with (S.41) and recalling that $D$ is independent of $h$, we obtain

$$\frac{\partial lnK_r}{\partial \hat{h}} = \frac{1}{2\Theta}\frac{\partial \Theta}{\partial \hat{h}} + \frac{2}{N}\frac{\partial N}{\partial \hat{h}}. \tag{S.45}$$

Now,

$$\frac{\partial N}{\partial \hat{h}} = \frac{\partial N}{\partial \Theta}\frac{\partial \Theta}{\partial \hat{h}} = \frac{\beta}{A\hat{h}} \cdot \frac{1}{\beta A\hat{h}^2(1+v^2)} = \frac{1}{A^2\hat{h}^3(1+v^2)}. \tag{S.46}$$

Substituting the expressions for $\partial\Theta/\partial$h and $\partial N/\partial$h gives

$$\frac{\partial lnK_r}{\partial \hat{h}} = \frac{1}{2\Theta}\frac{1}{\beta A\hat{h}^2(1+v^2)} + \frac{2}{N}\frac{1}{A^2\hat{h}^3(1+v^2)}. \tag{S.47}$$

A compact equivalent form is

$$S_{\mathrm{h}} := \frac{\partial lnK_r}{\partial \hat{h}} = \frac{1}{A\hat{h}^2(1+v^2)}\left[\frac{1}{2\beta\Theta} + \frac{2}{A\hat{h}N}\right]. \tag{S.48}$$

Similarly,

$$\frac{\partial lnK_r}{\partial A} = \frac{1}{2\Theta}\frac{\partial \Theta}{\partial A} + \frac{2}{N}\frac{\partial N}{\partial A}, \tag{S.49}$$

because $D$ does not depend on $A$. Using

$$\frac{\partial N}{\partial A} = \frac{\partial N}{\partial \Theta}\frac{\partial \Theta}{\partial A} = \frac{\beta}{A\hat{h}} \cdot \frac{1}{\beta A^2\hat{h}(1+v^2)} = \frac{1}{A^3\hat{h}^2(1+v^2)}, \tag{S.50}$$

we obtain

$$\frac{\partial lnK_r}{\partial A} = \frac{1}{2\Theta}\frac{1}{\beta A^2\hat{h}(1+v^2)} + \frac{2}{N}\frac{1}{A^3\hat{h}^2(1+v^2)}. \tag{S.51}$$

Thus

$$S_A := \frac{\partial lnK_r}{\partial A} = \frac{1}{A^2\hat{h}(1+v^2)}\left[\frac{1}{2\beta\Theta} + \frac{2}{A\hat{h}N}\right]. \tag{S.52}$$

The derivative with respect to $\beta$ is slightly more delicate because both $N$ and $D$ contain $\beta$ explicitly. We write

$$\frac{\partial lnK_r}{\partial \beta} = \frac{1}{2\Theta}\frac{\partial \Theta}{\partial \beta} + \frac{2}{N}\frac{\partial N}{\partial \beta} - \frac{2}{D}\frac{\partial D}{\partial \beta}. \tag{S.53}$$

Combined with (S.36) and (S.26), we can get the equivalent representation

$$N(h,\beta) = -\frac{1}{2}\ln(1+v^2) - ln|cos(\beta(\Theta_{low}+1))| - arctan v + \beta(1+\Theta_{low}) \tag{S.54}$$

At fixed $h$, the quantity $v$ is independent of $\beta$, and therefore

$$\frac{\partial N}{\partial \beta}=(1+\Theta_{low})tan\big(\beta(\Theta_{low}+1)\big)+(1+\Theta_{low}). \tag{S.55}$$

For $D$, direct differentiation gives

$$\frac{\partial D}{\partial \beta}=-\left(1+\Theta_{up}\right)\tan\left(\beta\left(\Theta_{up}+1\right)\right)+\left(1+\Theta_{low}\right)\tan\left(\beta\left(\Theta_{low}+1\right)\right)-\left(\Theta_{up}-\Theta_{low}\right) \tag{S.56}$$

Substituting these ingredients into the derivative of $\ln K_r$ yields

$$S_\beta:=\frac{\partial lnK_r}{\partial \beta}=-\frac{\Theta+1}{2\beta\Theta}+\frac{2}{N}[(1+\Theta_{low})tan(\beta(\Theta_{low}+1))+(1+\Theta_{low})]-\frac{2}{D}\frac{\partial D}{\partial \beta}. \tag{S.57}$$

Therefore, the first-order constitutive sensitivities for $K_r$ is:

$$\begin{gathered}S_h=\frac{1}{A\hat{h}^2(1+\nu^2)}\left[\frac{1}{2\beta\Theta}+\frac{2}{A\hat{h}N}\right],\\ S_A=\frac{1}{A^2\hat{h}(1+\nu^2)}\left[\frac{1}{2\beta\Theta}+\frac{2}{A\hat{h}N}\right],\\ S_\beta=-\frac{\Theta+1}{2\beta\Theta}+\frac{2}{N}[(1+\Theta_{low})tan(\beta(\Theta_{low}+1))+(1+\Theta_{low})]-\frac{2}{D}\frac{\partial D}{\partial \beta}.\end{gathered} \tag{S.58}$$

These are exactly the coefficients needed to construct a first-order stochastic perturbation model (Zhang, 2002).

**(2) First-order constitutive sensitivities for $\theta$ and $K_r$ when $p$=3**

When $p$=3, the constitutive law is written as:

$$1/\hat{h}=A[-\tan^3(\beta(\Theta+1))-1] \tag{S.59}$$

We first introduce the same auxiliary variable as in the $p$=1 case, together with its real cubic root. This notation keeps the inversion and the derivative expressions compact.

$$\begin{gathered}\nu(\hat{h};A):=-\frac{1}{A\hat{h}}-1,\\ \omega(\hat{h};A):=\nu^{1/3}.\end{gathered} \tag{S.60}$$

Then the retention relation becomes $tan^3(\beta(\Theta+1))=\nu$, or equivalently $tan(\beta(\Theta+1))=\omega$ on the chosen real branch. Therefore, we have

$$\begin{gathered}\beta(\Theta+1)=\arctan(\omega),\\ \Theta(\hat{h};A,\beta)=\frac{1}{\beta}\arctan(\omega)-1.\end{gathered} \tag{S.61}$$

This again provides an explicit inversion from $h$ to Θ. The difference from the $p$=1 case is that the argument of arctan is no longer ν itself, but its real cubic root $\omega=\nu^{1/3}$.

We now differentiate Θ with respect to $h$. Since $\omega=\nu^{1/3}$ in the real sense, the chain rule gives:

$$\begin{gathered}\frac{\partial \omega}{\partial \hat{h}}=\frac{1}{3}\omega^{-2}\frac{\partial \nu}{\partial \hat{h}},\quad \frac{\partial \nu}{\partial \hat{h}}=\frac{1}{A\hat{h}^2}.\\ \frac{\partial \Theta}{\partial \hat{h}}=\frac{1}{\beta}\cdot\frac{1}{1+\omega^2}\cdot\frac{\partial \omega}{\partial \hat{h}}=\frac{1}{3\beta A\hat{h}^2\omega^2(1+\omega^2)}.\end{gathered} \tag{S.62}$$

Therefore, we obtain:

$$C_{\hat{h}}:=\frac{\partial \theta}{\partial \hat{h}}=\frac{\theta_s-\theta_r}{3\beta A\hat{h}^2\omega^2(1+\omega^2)}. \tag{S.63}$$

Similarly, differentiation with respect to A yields:

$$\frac{\partial \nu}{\partial A}=\frac{1}{A^2\hat{h}}, \tag{S.64}$$

$$\frac{\partial \omega}{\partial A}=\frac{1}{3}\omega^{-2}\frac{\partial \nu}{\partial A}=\frac{1}{3A^2\hat{h}\omega^2}.$$

$$\frac{\partial \Theta}{\partial A}=\frac{1}{\beta}\cdot\frac{1}{1+\omega^2}\cdot\frac{\partial \omega}{\partial A}=\frac{1}{3\beta A^2\hat{h}\omega^2(1+\omega^2)}.$$

Because $\nu$ and $\omega$ do not depend on $\beta$, the derivative with respect to $\beta$ is unchanged in structure relative to the $p$=1 case:

$$C_\beta:=\frac{\partial \theta}{\partial \beta}=-(\theta_s-\theta_r)\frac{\Theta+1}{\beta}. \tag{S.65}$$

For $p$=3, the Mualem-type integral in the numerator can still be evaluated analytically. Let $u$=$\beta(\chi + 1)$, we have

$$\int -\tan^3(u)\,du=-\frac{1}{2}\tan^2(u)-\ln\cos u.$$

$$\int_{\Theta_{\mathrm{low}}}^{\Theta}\frac{1}{\hat{h}(\chi)}d\chi=\frac{A}{\beta}N(\Theta), \tag{S.66}$$

$$N(\Theta):=-\frac{1}{2}\left[\tan^2(\beta(\Theta+1))-\tan^2(\beta(\Theta_{\mathrm{low}}+1))\right]-\ln\left[\frac{\cos(\beta(\Theta+1))}{\cos(\beta(\Theta_{\mathrm{low}}+1))}\right]-\beta(\Theta-\Theta_{\mathrm{low}}).$$

Similarly, we define:

$$D:=-\frac{1}{2}\left[\tan^2\big(\beta(\Theta_{up}+1)\big)-\tan^2(\beta(\Theta_{low}+1))\right]-\ln\left[\frac{\cos\big(\beta(\Theta_{up}+1)\big)}{\cos(\beta(\Theta_{low}+1))}\right]-\beta(\Theta_{up}-\Theta_{low}). \tag{S.67}$$

Then the relative conductivity can be written as:

$$K_r(\Theta)=\Theta^{1/2}\left[\frac{N(\Theta)}{D}\right]^2,$$
$$\ln K_r=\frac{1}{2}\ln\Theta+2\ln|N|-2\ln|D|. \tag{S.68}$$

Differentiating Eq. (S.66) gives:

$$\frac{\partial N}{\partial \Theta}=-\beta\tan(\beta(\Theta+1))\sec^2(\beta(\Theta+1))+\beta\tan(\beta(\Theta+1))-\beta. \tag{S.69}$$

Now we use

$$\tan(\beta(\Theta+1))=\omega,\qquad \sec^2(\beta(\Theta+1))=1+\omega^2. \tag{S.70}$$

Then the first two terms become

$$-\beta\omega(1+\omega^2)+\beta\omega=-\beta\omega^3. \tag{S.71}$$

Since $\omega^3$=$\nu$=-1/($A\hat{h}$)-1, we obtain

$$\frac{\partial N}{\partial \Theta}=-\beta\omega^3-\beta=\frac{\beta}{A\hat{h}}. \tag{S.72}$$

Because $D$ is independent of $h$ and $A$ when $\Theta_{\mathrm{low}}$, $\Theta_{\mathrm{up}}$, and $\beta$ are fixed, the derivatives of ln$K_r$ with respect to $\hat{h}$ and $A$ are again simpler than the derivatives of $K_r$ itself. We have

$$\frac{\partial \ln K_r}{\partial \hat{h}}=\frac{1}{2\Theta}\frac{\partial \Theta}{\partial \hat{h}}+\frac{2}{N}\frac{\partial N}{\partial \hat{h}}.$$
$$\frac{\partial N}{\partial A}=\frac{\partial N}{\partial \Theta}\frac{\partial \Theta}{\partial A}=\frac{\beta}{A\hat{h}}\cdot\frac{1}{3\beta A^2\hat{h}\omega^2(1+\omega^2)}=\frac{1}{3A^3\hat{h}^2\omega^2(1+\omega^2)}. \tag{S.73}$$

Hence

$$S_A:=\frac{\partial \ln K_r}{\partial A}=\frac{1}{3A^2\hat{h}\omega^2(1+\omega^2)}\left[\frac{1}{2\beta\Theta}+\frac{2}{A\hat{h}N}\right]. \tag{S.74}$$

The derivative with respect to $\beta$ is again more delicate because both $N$ and $D$ depend explicitly on $\beta$. At fixed $\hat{h}$, the quantity $\omega$ is independent of $\beta$, and $N$ can be rewritten as:

$$N(\hat{h},\beta)=-\frac{1}{2}\omega^2+\frac{1}{2}\tan^2(\beta(\Theta_{\mathrm{low}}+1))+\frac{1}{2}\ln(1+\omega^2) \\ +\ln\cos(\beta(\Theta_{\mathrm{low}}+1))-\arctan(\omega)+\beta(\Theta_{\mathrm{low}}+1) \tag{S.75}$$

Similarly,

$$\frac{\partial D}{\partial \beta}=(1+\Theta_{low})\left[\tan^3(\beta(\Theta_{low}+1))+1\right]-(1+\Theta_{up})\left[\tan^3\left(\beta(\Theta_{up}+1)\right)+1\right] \tag{S.76}$$

Thus

$$\frac{\partial \ln K_r}{\partial \beta}=\frac{1}{2\Theta}\frac{\partial \Theta}{\partial \beta}+\frac{2}{N}\frac{\partial N}{\partial \beta}-\frac{2}{D}\frac{\partial D}{\partial \beta}. \tag{S.77}$$

Using $\frac{\partial\Theta}{\partial\beta}=-\frac{\Theta+1}{\beta}$, we have:

$$S_\beta:\ =\frac{\partial \ln K_r}{\partial \beta}=-\frac{\Theta+1}{2\beta\Theta}+\frac{2}{N}(1+\Theta_{low})\left[\tan^3(\beta(\Theta_{low}+1))+1\right] \\ -\frac{2}{D}\left\{(1+\Theta_{\mathrm{low}})\left[\tan^3(\beta(\Theta_{\mathrm{low}}+1))+1\right]-(1+\Theta_{\mathrm{up}})\left[\tan^3\left(\beta(\Theta_{\mathrm{up}}+1)\right)+1\right)\right. \tag{S.78}$$

## Text S6

Under the capillary-bundle assumption, the pore radius satisfies $r = C/h$, where $h>0$ is the suction head and $C$ is the capillary constant. The discovered retention function therefore implies the pore-volume density

$$f_r(r) = -\frac{C}{r^2}\frac{d\Theta}{dh}\bigg|_{h=C/r} = \frac{(1+r/r_*)^{1/p-1}}{bp\,r_*[1+(1+r/r_*)^{2/p}]}, \tag{S.79}$$

where $b = -\beta > 0$, $r_* = AC/h_0$, and $h_0$ is the reference head. For $p \in \{1,3\}$, this density is positive and decreases with radius, whereas the density per unit logarithmic radius, $rf_r(r)$, has a single interior maximum on the complete branch. Its endpoint behaviors are

$$\lim_{r\to 0^+} f_r(r) = \frac{1}{2bp\,r_*}, \qquad f_r(r) \propto r^{-1-1/p} \quad (r\to\infty). \tag{S.80}$$

The resulting large-pore tail contrasts with the finite upper cutoff of Brooks–Corey and the faster decay of standard MvG and Kosugi distributions. Complete-branch normalization requires $b = \pi/4$, since

$$\int_0^\infty f_r(r)\,dr = \frac{\pi}{4b}. \tag{S.81}$$

Over a finite admissible interval, the corresponding conditional density is

$$f_{\mathrm{adm}}(r) = \frac{f_r(r)}{\Theta_{\mathrm{up}}-\Theta_{\mathrm{low}}}, \qquad r_{\mathrm{low}} \le r \le r_{\mathrm{up}}. \tag{S.82}$$

The complete-branch first moment diverges for either exponent, making the finite wet-end bound relevant to the conductivity integral. These properties provide an equivalent capillary pore-volume interpretation within the specified domain, with $r_*$ setting the pore scale and $p$ controlling the large-pore tail.

## Text S7

We performed model discovery directly on the UNSODA dataset, and the resulting candidate functions (Table S2) again showed a clear preference for the trigonometric family. However, the performance of their retention and conductivity differed. Rank 1 yielded a higher retention RMSE than the original model (0.046 versus 0.021), while its conductivity RMSE was slightly lower (2.839 versus 2.862). Rank 2 yielded higher retention and conductivity errors (0.064 and 3.069, respectively). Rank 3 failed the analytical-integrability criterion adopted here and was excluded from prediction evaluation. For the present model and the rank-1 to rank-3 tangent-form models, fitting to the UNSODA soils on the principal branch yields no singularities within the range of fitted pressure heads. Ranks 4 and 5 were refitted using the same retention-only regression protocol, with parameter constraints placing their singularities outside the fitted saturation range. Both yielded a retention RMSE of 0.045, while their conductivity RMSEs were 2.669 and 2.783, respectively. These conductivity errors were lower than that of the original model (2.862), but higher than that of the modified model. This result underscores that equation discovery is constrained not only by data quantity, but also by data quality and informativeness. UNSODA usually contains measurements with uneven quality and incomplete pressure-head coverage, which can obscure the constitutive structure to be discovered. In addition, our search is performed in the transformed space $1/h=f(\theta)$ to preserve analytical integrability for calculating $K_r$, so a lower discovery loss does not necessarily imply lower errors in $\theta$ or $K_r$ after inversion and integration.

**Text S8**

In the main text, we compared the discovered tangent model with classical Mualem–van Genuchten (MvG), an air-entry modification of MvG, bimodal MvG (Durner), and Peters–Durner–Iden MvG (PDI-MvG). For each soil, calibration used the measured water retention curve. Relative hydraulic conductivity was then predicted using the prescribed Mualem or PDI relationship. Conductivity observations were reserved for evaluation. All soil-specific choices, including the tangent exponent and the air-entry head, were determined from retention data.

We denote the dimensionless magnitude of pressure head by $\hat{h}$ and volumetric water content by $\theta$. For models with a conventional residual water content, effective saturation is $\Theta$. The saturated water content $\theta_s$ was fixed from the UNSODA file. Measured saturated hydraulic conductivity $K_s$ supplied the reference for observed relative conductivity, $K_r^{\mathrm{obs}} = K^{\mathrm{obs}}/\mathrm{K}_s$. The pore-connectivity exponent was fixed at $L = 1/2$ throughout the comparison. Conductivity measurements did not enter any stage of parameter or structural selection. Thus, the shared calibration protocol comprised retention-only estimation, with model-specific bounds and optimization procedures. Conductivity error used the natural logarithm, and a common evaluation floor of $\varepsilon = 10^{-8}$. For each soil, the error was

$$\mathrm{RMSE}_{\ln K_r} = \left[\frac{1}{N_K}\sum_{j=1}^{N_K}\left\{\ln\left[\max\left(K_{r,j}^{\mathrm{pred}}, \varepsilon\right)\right] - \ln\left[\max\left(K_{r,j}^{\mathrm{obs}}, \varepsilon\right)\right]\right\}^2\right]^{1/2}. \tag{S.83}$$

The floor made zero or very small relative conductivities evaluable in logarithmic space. A failed fit therefore remained in the comparison through its fallback output, with zero soil exclusions based solely on optimizer failure. Details for these models are provided below.

1. Classical Mualem-van Genuchten model

The classical MvG model used

$$S_e(h) = [1 + (\alpha h)^n]^{-m}, \quad m = 1 - \frac{1}{n} \tag{S.84}$$

$$\theta(h) = \theta_r + (\theta_s - \theta_r)S_e(h).$$
$$K_r(h) = S_e(h)^{1/2}\left[1 - \left(1 - S_e(h)^{1/m}\right)^m\right]^2$$

We fitted $\theta_r$, $\alpha$, and $n$ using curve_fit, with a maximum of 80,000 function evaluations. The bounds were $10^{-8} < \theta_r < \min_j \theta_j^{obs}$, $10^{-8} \le \alpha \le 10^2$, and $1.01 \le n \le 10$. The selected fit minimized retention mean squared error.

2. Air entry modification of the Mualem-van Genuchten model

The air-entry modification imposed a saturated plateau below a selected entry head $h_a$. Defining $G(h) = [1 + (\alpha h)^n]^{-m}$ with $m = 1 - 1/n$, the effective saturation was

$$S_e(h) = \begin{cases} 1, & h \le h_a, \\ \dfrac{G(h)}{G(h_a)}, & h > h_a, \end{cases} \tag{S.85}$$

Water content was calculated as $\theta(h) = \theta_r + (\theta_s - \theta_r)S_e(h)$. For each candidate $h_a \in \{1,2,5,10,25\}$ cm, the continuous parameters $\theta_r$, $\alpha$, and $n$ were fitted using the bounds and optimization framework of classical MvG. The combination with the smallest retention mean squared error was retained. If the modified fit failed, we selected the first entry-head candidate, $h_a = 1$ cm. The conductivity calculation used numerical Mualem integration with 1,600 points in effective saturation. For $0 < S_e < 1$, the inverse retention relation entering that integration was

$$h(S_e) = \frac{1}{\alpha}\left[\left(S_e G(h_a)\right)^{-1/m} - 1\right]^{1/n} \tag{S.86}$$

The corresponding Mualem expression was

$$K_r(S_e) = S_e^{1/2}\left[\frac{\int_0^{S_e} \frac{1}{h(u)}\,du}{\int_0^1 \frac{1}{h(u)}\,du}\right]^2 \tag{S.87}$$

The implementation assigned $K_r = 1$ at $S_e = 1$ and clipped final predictions to $[0,1]$. Selection of $h_a$ used retention measurements, so it constituted one discrete fitted choice in addition to the three continuous coefficients.

3. Bimodal Mualem-van Genuchten model

The Durner model represented effective saturation as:

$$S_e(h) = wS_{e1}(h) + (1 - w)S_{e2}(h).$$
$$S_{ei}(h) = [1 + (\alpha_i h)^{n_i}]^{-m_i}, \quad m_i = 1 - \frac{1}{n_i}, \quad i \in \{1,2\}. \tag{S.88}$$

With $w_1 = w$ and $w_2 = 1 - w$, the analytical bimodal Mualem prediction was

$$K_r(h) = S_e(h)^{1/2}\left[\frac{\sum_{i=1}^{2} w_i\, \alpha_i \left\{1 - \left[1 - S_{ei}(h)^{1/m_i}\right]^{m_i}\right\}}{\sum_{i=1}^{2} w_i\, \alpha_i}\right]^2 \tag{S.89}$$

We fitted six continuous coefficients, $\theta_r$, $w$, $\alpha_1$, $n_1$, $\alpha_2$, and $n_2$, using five initial parameter sets and curve_fit with a maximum of 30,000 function evaluations. The bounds were $10^{-8} < \theta_r < \min\left(\min(\theta^{obs}), \theta_s\right)$, $10^{-4} \le w \le 1 - 10^{-4}$, $10^{-8} \le \alpha_i \le 10^2$, and $1.01 \le n_i \le 10$. Starting mixture weights ranged from 0.3 to 0.9. The best fit was selected using retention mean squared error. After fitting, the components were ordered so that $\alpha_1 \le \alpha_2$. If necessary, the component parameters were exchanged and the weight was replaced by $1 - w$.

4. Peters Durner Iden (PDI) Mualem-van Genuchten model

PDI-MvG model can be expressed as:

$$\Gamma(h) = [1 + (\alpha h)^n]^{-m}, \quad m = 1 - \frac{1}{n}.$$
$$S_c(h) = \frac{\Gamma(h) - \Gamma(h_0)}{1 - \Gamma(h_0)}. \tag{S.90}$$

The capillary branch was truncated at the prescribed dry head $h_0$. Total water content combined capillary retention with the PDI noncapillary saturation function,

$$\theta(h) = (\theta_s - \theta_{r,\mathrm{nc}})S_c(h) + \theta_{r,\mathrm{nc}}S_{\mathrm{nc}}(h). \tag{S.91}$$

Here $S_{\mathrm{nc}}(h)$ is the adsorption or film-retention function. The fitted coefficient $\theta_{r,\mathrm{nc}}$ scales this contribution. Its interpretation differs from the conventional residual water content used in the other four models, and its upper bound could exceed the minimum observed water content. Five initial parameter sets were evaluated using curve_fit with a maximum of 30,000 function evaluations. The retained set minimized retention mean squared error. Conductivity was calculated by the PDI module using the fitted retention coefficients and fixed hydraulic options. The settings were hclip=6, afilm=1.5, and TC=20 degrees Celsius. These module options defined the conductivity calculation, while all three fitted coefficients came from retention observations. The resulting prediction was clipped to $[10^{-12}, 1]$.

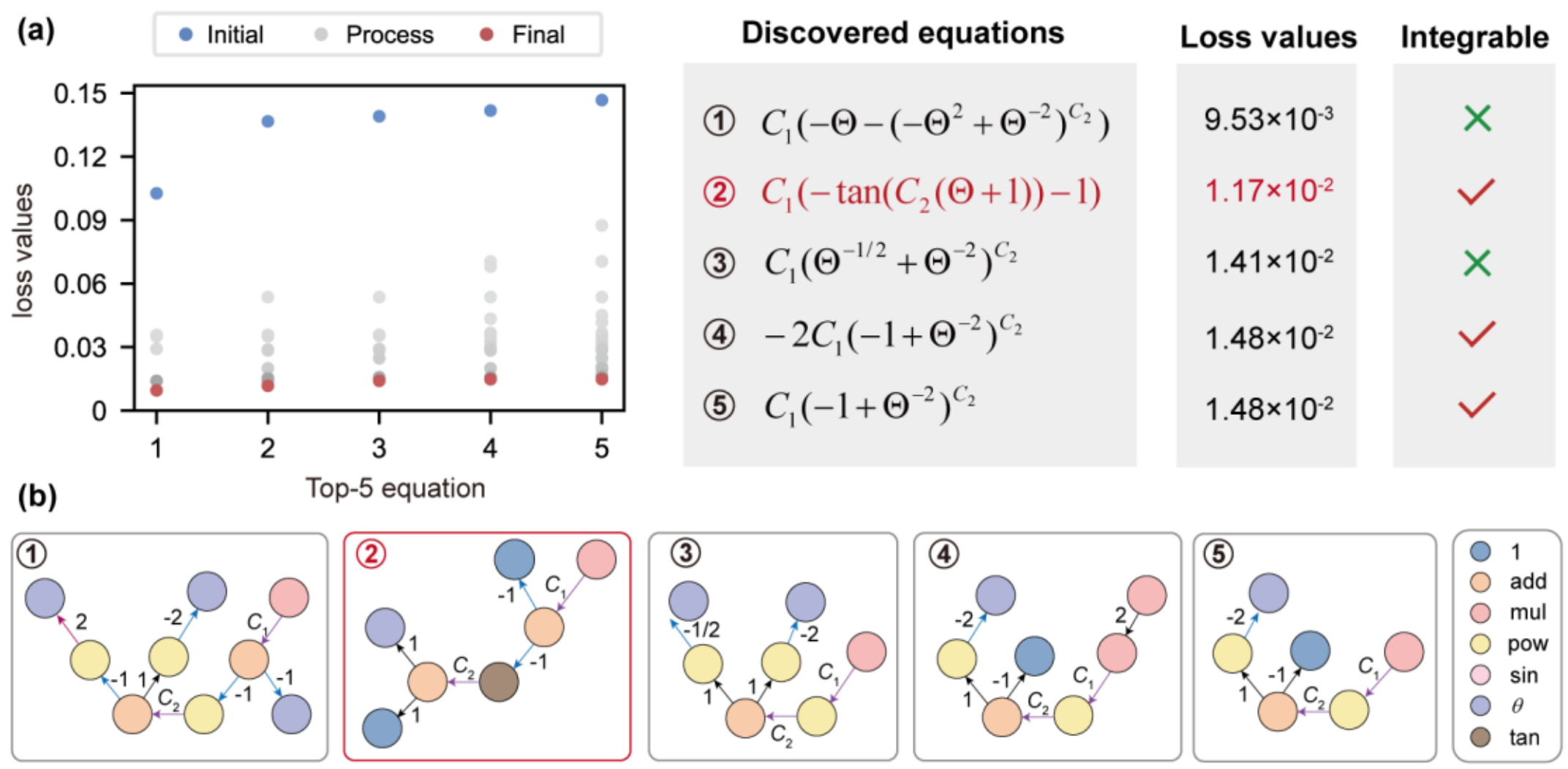

**Figure S1. Candidate equations identified during graph-based model discovery.** (a) Optimization trajectories, mathematical expressions, loss values, and analytical integrability of the five highest-ranked equations. The selected tangent-based equation is highlighted in red. (b) Corresponding graph representations of the five candidate equations.

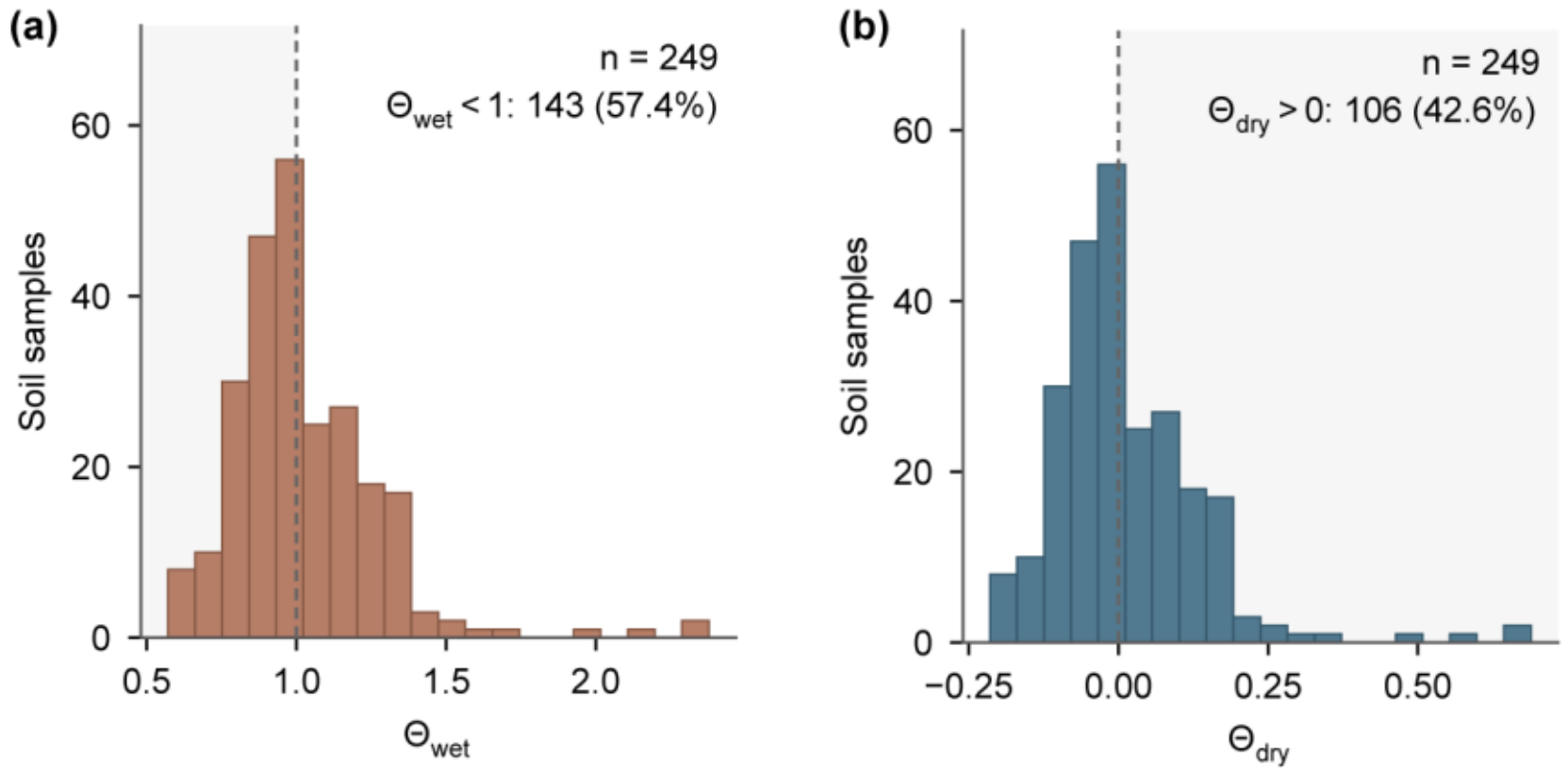

**Figure S2. Distribution of $\Theta_{wet}$ and $\Theta_{dry}$ across 249 soil samples.** Shading indicates the proportion of samples within the physically admissible range.

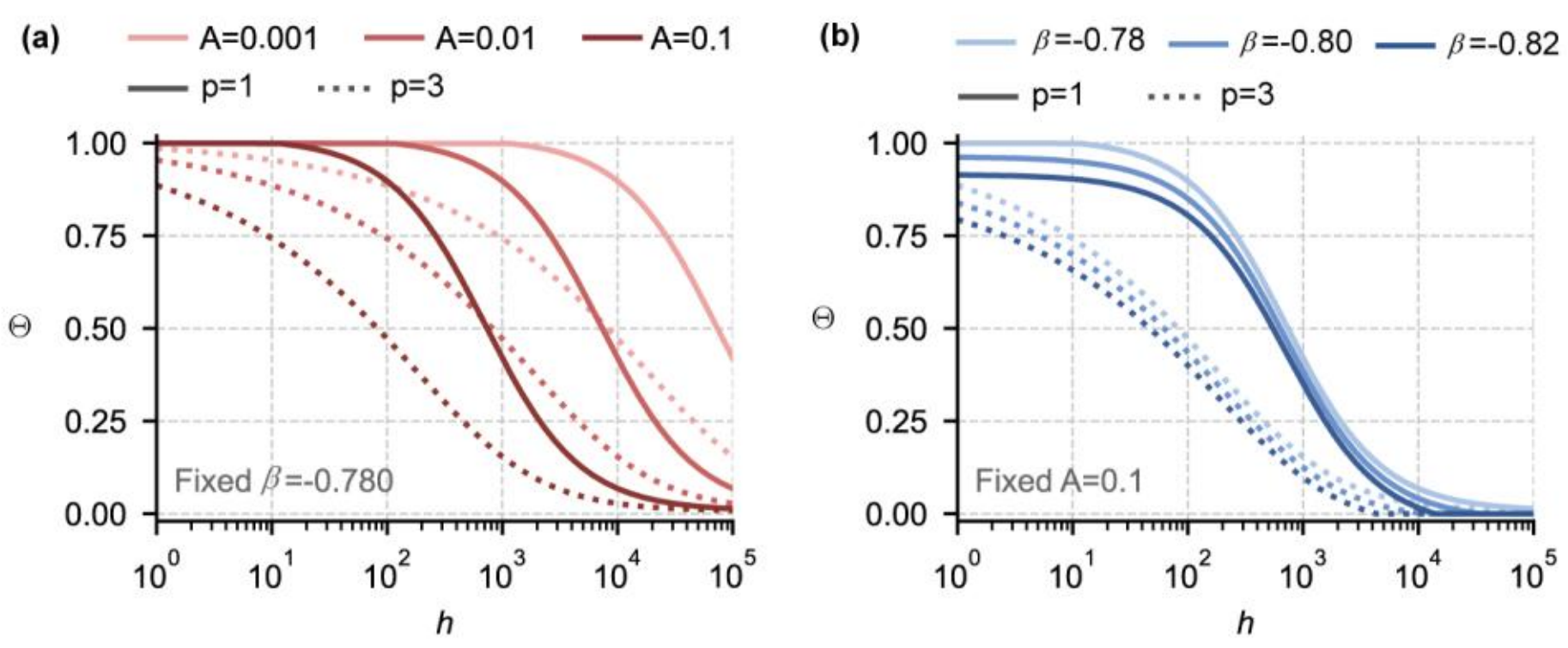

**Figure S3. Effects of model parameters on the normalized soil water retention curves of the modified model.** (a) Influence of $A$ on the soil water retention curves with $\beta=-0.780$ fixed. (b) Influence of $\beta$ with $A=0.1$ fixed. Solid and dotted lines denote $p=1$ and $p=3$, respectively.

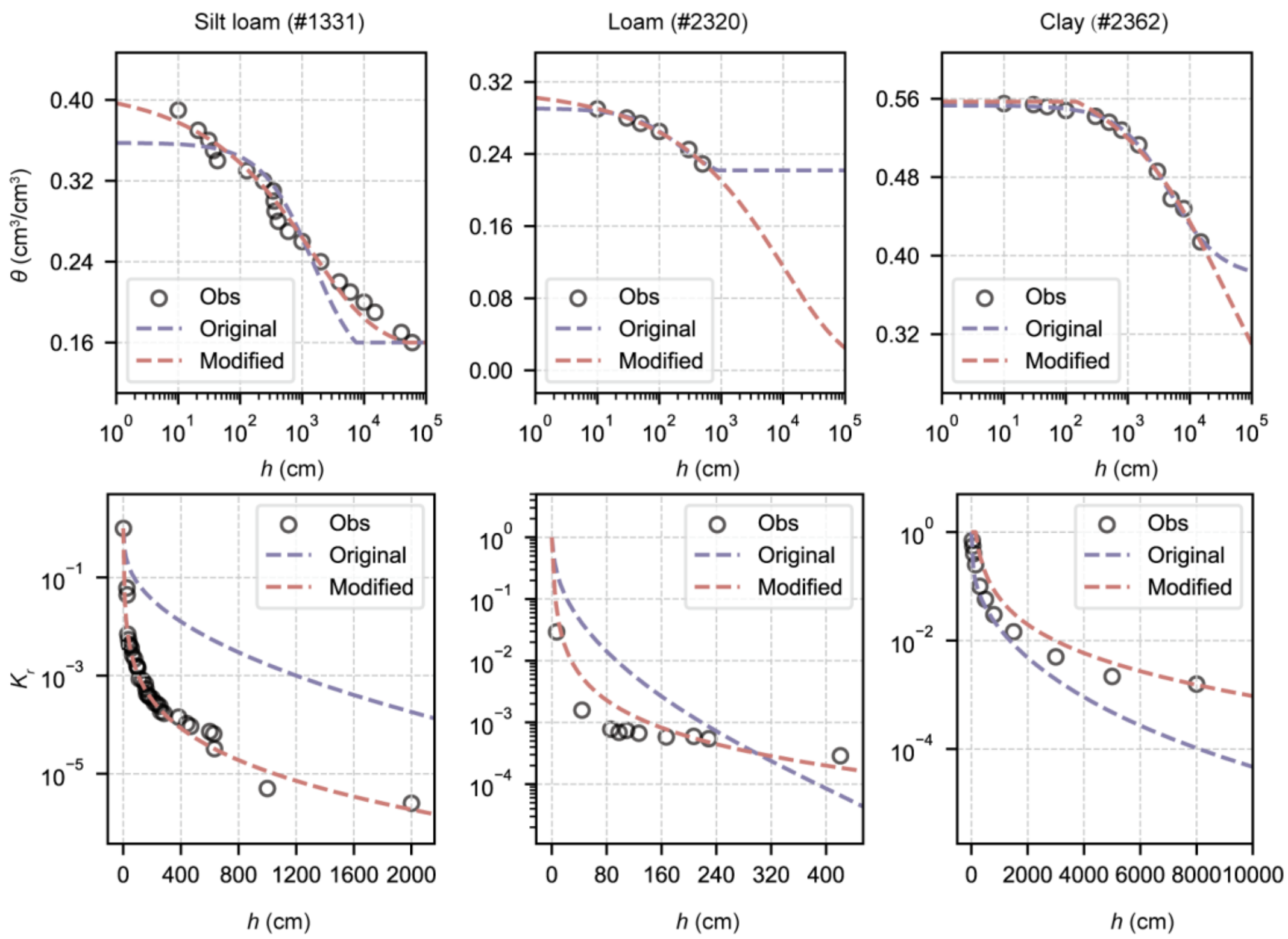

**Figure S4. Representative comparisons between the original and modified discovered models.** Observed and predicted soil water retention curves are shown in the upper row, with the corresponding relative hydraulic conductivity curves in the lower row, for silt loam, loam, and clay soils.

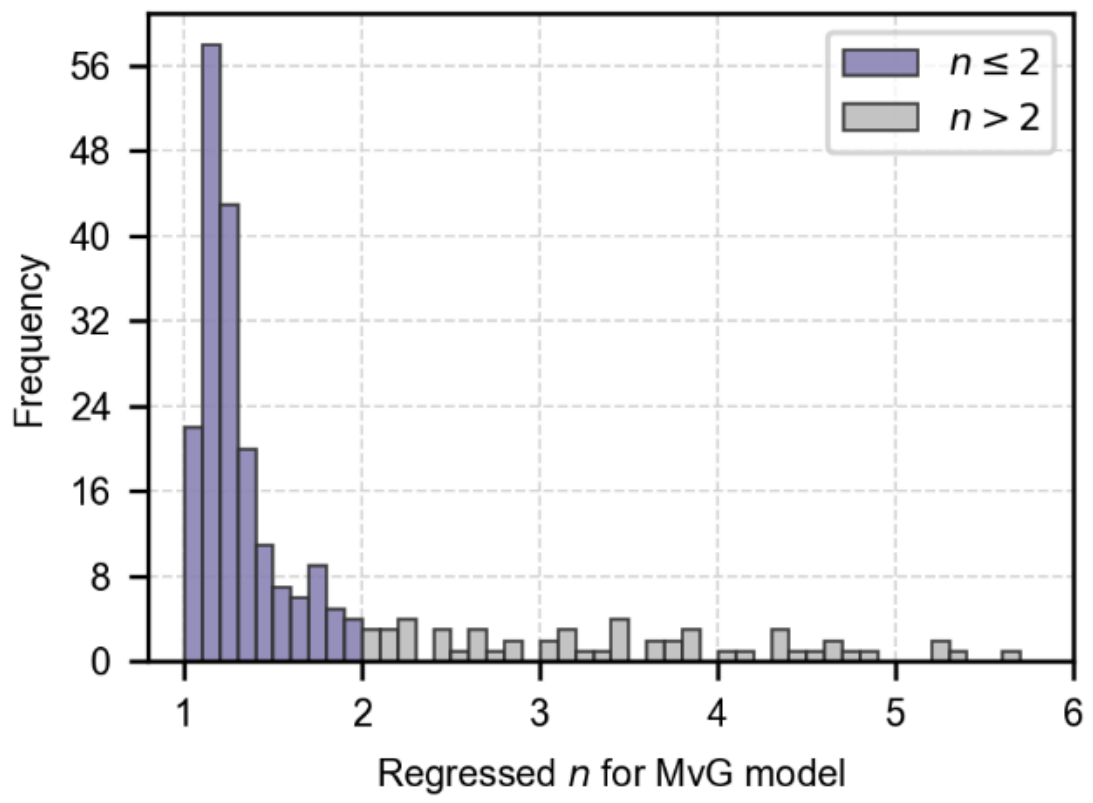

**Figure S5. Histogram of regressed *n* values for the Mualem–van Genuchten (MvG) model.** The bars represent the frequency distribution of the fitted *n* values in the MvG model, which are categorized into two groups: $n \le 2$ (purple) and $n > 2$ (grey).

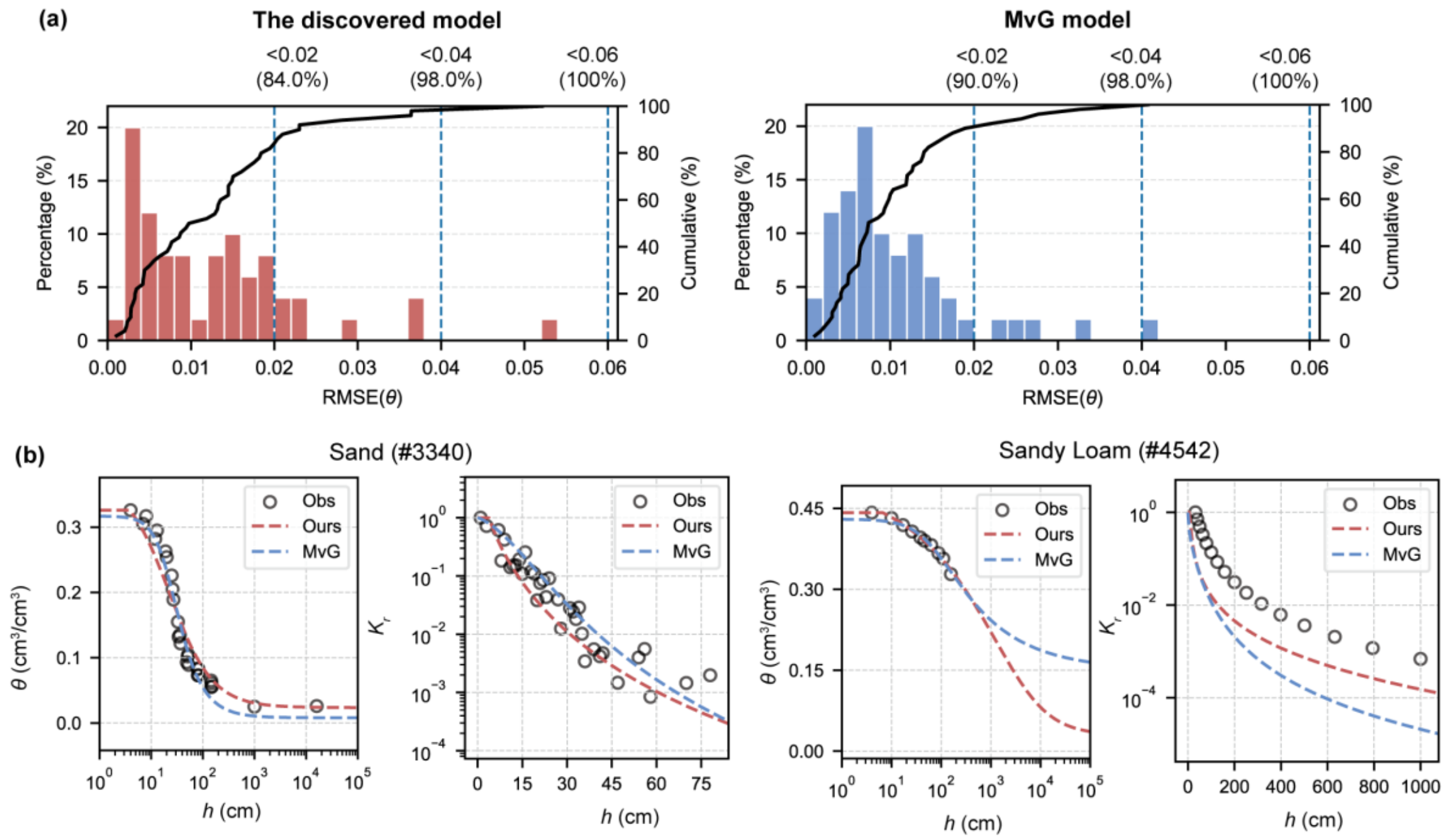

**Figure S6. Comparison of soil water retention performance between the discovered model and the Mualem–van Genuchten model.** (a) Distributions and cumulative percentages of retention fitting errors across soils on the validation set of UNSODA. (b) Representative retention and hydraulic conductivity predictions for sand and sandy loam soils.

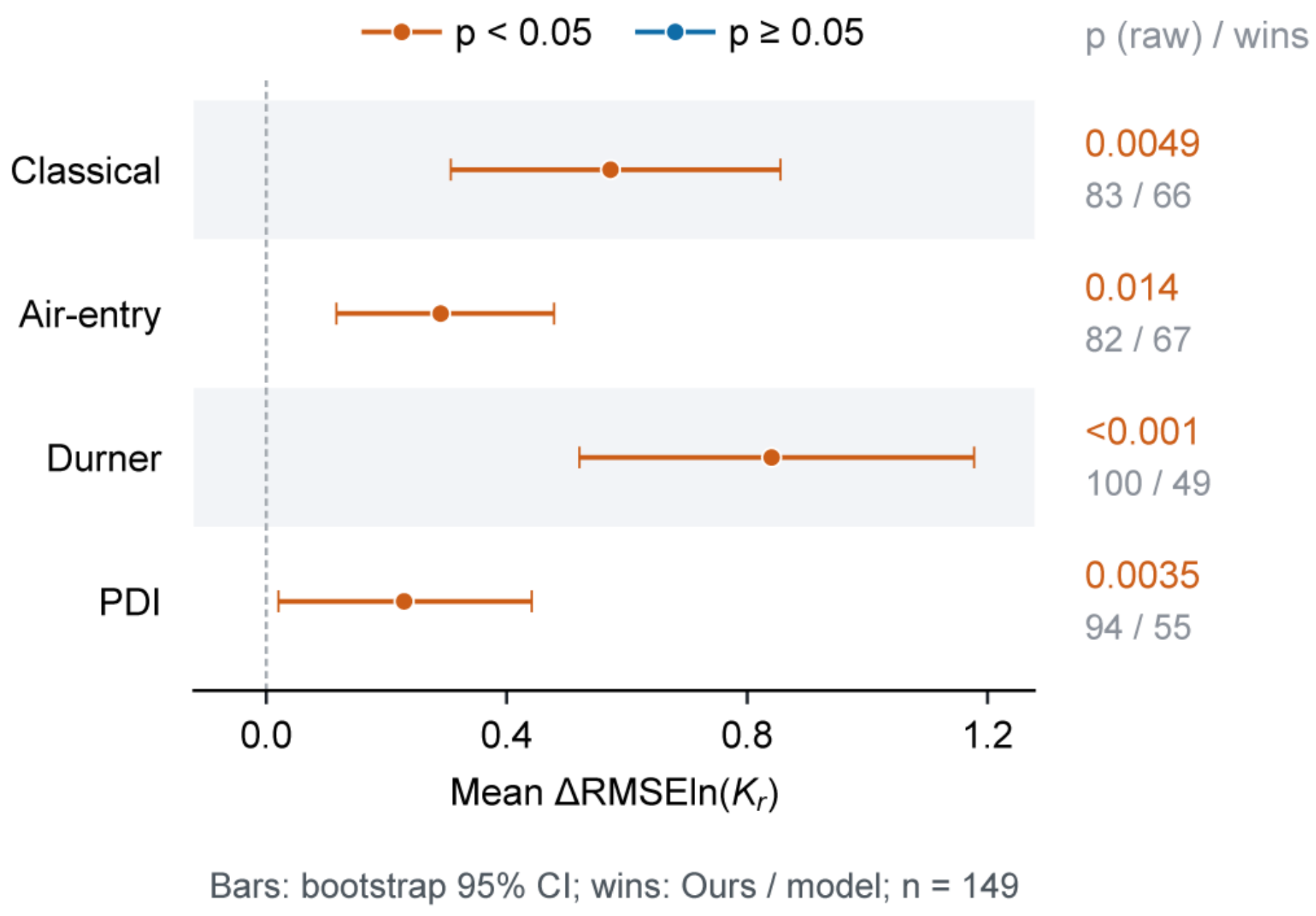

**Figure S7. Mean ΔRMSE in ln($K_r$) of the discovered model relative to classical models.** Bootstrap 95% confidence intervals and paired win counts are shown. $p$ denotes the raw two-sided Wilcoxon $p$-value. ΔRMSE is the difference between the discovered model and compared model, and positive values favor the discovered model.

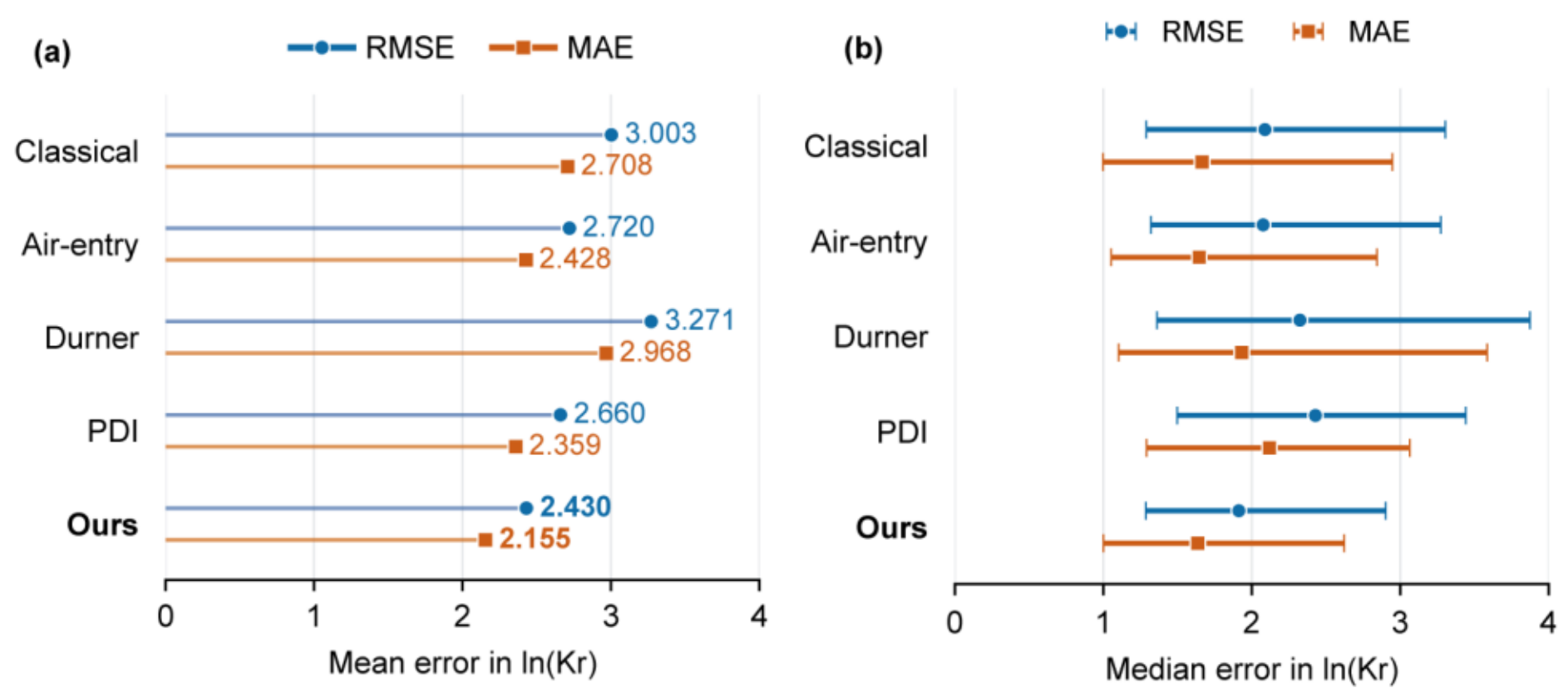

**Figure S8. Mean and median errors in ln($K_r$) for the discovered model and classical formulations.** (a) Mean RMSE and MAE. (b) Median RMSE and MAE. In (b), horizontal bars span the interquartile range (Q25 – Q75).

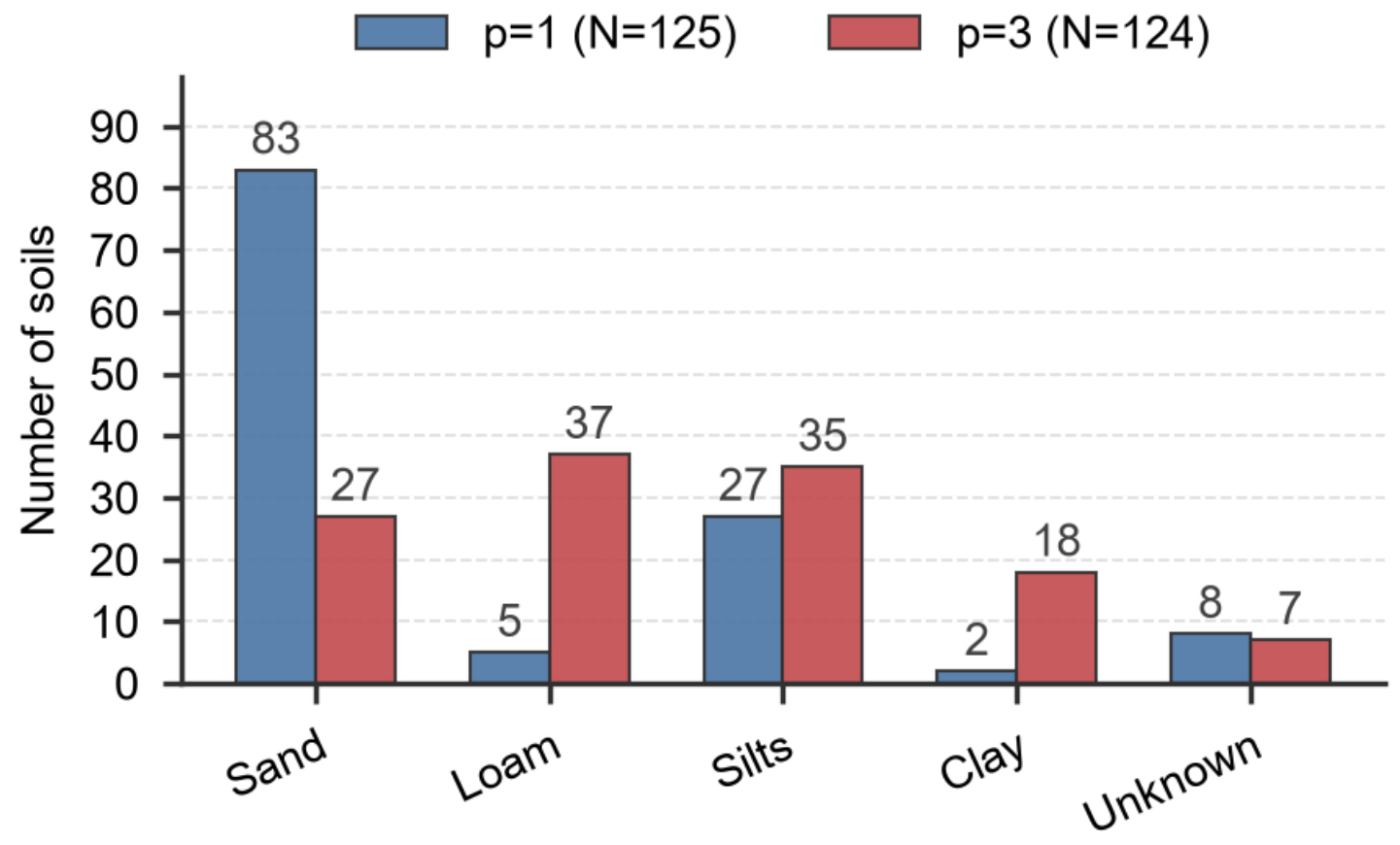

**Figure S9. Distribution of the optimized parameter *p* across soil texture classes in the modified model.** Numbers above bars indicate sample counts.

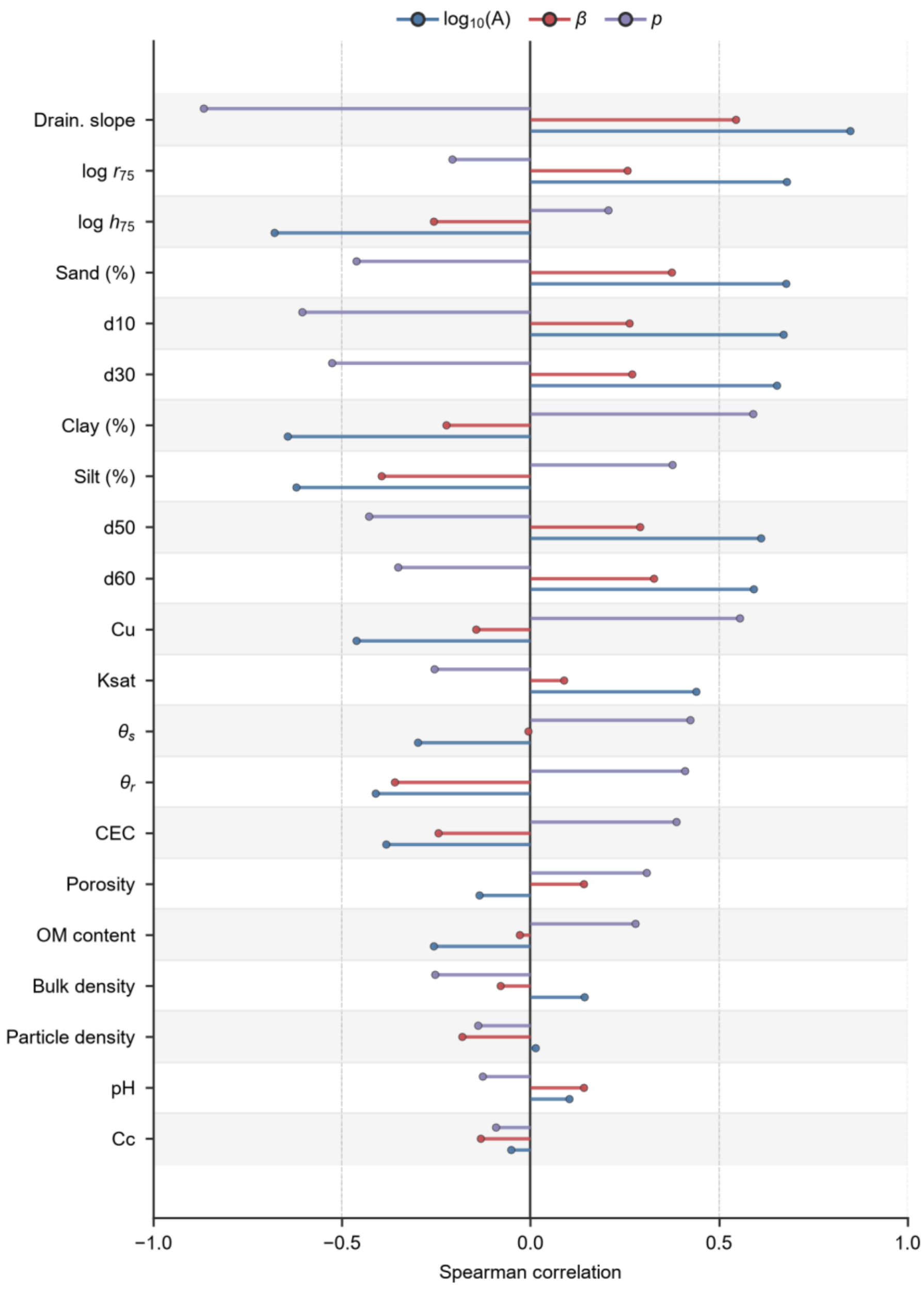

**Figure S10. Spearman correlations between fitted model parameters and soil physical descriptors.** The plot shows the correlations of $\log_{10}(A)$, $\beta$, and $p$ with drainage-curve metrics, pore-size indicators, particle-size fractions and basic soil properties. Positive and negative values indicate the direction of association, and the vertical line marks zero correlation.

**Table S1. Equations discovered from five independent random seeds and their corresponding fitness values.**

| Random seeds | Discovered equation | Fitness |
|---|---|---|
| 1 | $A(x - \tan(C_4(-x-1)) - 1)$ | $7.79 \times 10^{-3}$ |
| 2 | $A(\tan(C_2(-x-1)) - 1)$ | $1.17 \times 10^{-2}$ |
| 3 | $A(-\tan(C_4 x - 1) - 1)$ | $9.48 \times 10^{-3}$ |
| 4 | $A\dfrac{x^{-1/2} + x^2}{-x^{1/2} + x^{C_8}}$ | $6.21 \times 10^{-3}$ |
| 5 | $A(-\tan(C_3 x + \pi x - 1) - 1)$ | $9.48 \times 10^{-3}$ |

**Table S2. Summary of the top five discovered candidate functions from UNSODA data.** Integrable means whether the function is analytically integrable. RMSE$_\theta$ denotes the root mean squared error for soil water content, and RMSE$_{\ln(Kr)}$ denotes the root mean squared error for logarithmic relative hydraulic conductivity. Values in parentheses indicate the corresponding standard deviations across samples. For comparison, the performance metrics of the primary formulation Eq. (3) discovered in the main text are included in the first row as a benchmark.

| Rank | Discovered function | Loss | Integrable | RMSE$_\theta$ | RMSE$_{\ln(Kr)}$ |
|---|---|---|---|---|---|
| Eq. (3) | $C_1(-\tan(C_2(\Theta+1))-1)$ | 3.251 | √ | 0.021 (0.018) | 2.862 (1.749) |
| 1 | $C_1(-\Theta^{1/2}+\tan(C_2\Theta))$ | 2.037 | √ | 0.046 (0.041) | 2.839 (1.973) |
| 2 | $1/(\tan(C_1\Theta+1))^2$ | 2.086 | √ | 0.064 (0.059) | 3.069 (2.723) |
| 3 | $C_1\Theta(-\tan(C_2\Theta)-1)$ | 2.091 | × | × | × |
| 4 | $C_1\Theta/(C_2-\Theta)^2$ | 2.094 | √ | 0.045 (0.038) | 2.669 (1.926) |
| 5 | $-C_1\Theta^2/(C_2-\Theta)^2$ | 2.101 | √ | 0.045 (0.041) | 2.783 (2.341) |

**Table S3. Parameter structure of the discovered model and classical Mualem–van Genuchten (MvG) variants.** Continuous coefficients are estimated within the prescribed bounds, while discrete choices and dependent parameters follow the indicated relations or restricted sets.

| Model | Continuous fitted coefficients | Discrete fitted choice and parameter dependency |
|---|---|---|
| Discovered model | $A, \beta, \theta_r$ | $p \in \{1,3\}$ |
| Classical MvG | $\theta_r, \alpha, n$ | $m = 1 - 1/n$ |
| Air entry MvG | $\theta_r, \alpha, n$ | $h_a \in \{1,2,5,10,25\}$,<br>$m = 1 - 1/n$ |
| Bimodal MvG | $\theta_r, w, \alpha_1, n_1, \alpha_2, n_2$ | $m_1 = 1 - 1/n_1$<br>$m_2 = 1 - 1/n_2$ |
| PDI-MvG | $\theta_{r,nc}, \alpha, n$ | $m = 1 - 1/n$ |

**Table S4. Comparison of relative hydraulic conductivity prediction errors for the proposed model and the original and truncated MvG and PDI-MvG models.** Percentages indicate the proportion of soils below each RMSE threshold.

| **Models** | Classical MvG | Classical MvG with truncation | PDI-MvG | PDI-MvG with truncation | Ours |
|---|---|---|---|---|---|
| **$RMSE_{ln(Kr)}$** | 3.003 | 2.811 | 2.660 | 2.639 | **2.430** |
| **$RMSE_{ln(Kr)}$<2.5** | 61.1% | 60.4% | 52.3% | 51.7% | **65.8%** |
| **$RMSE_{ln(Kr)}$<5** | 84.6% | 88.6% | 92.6% | **94.0%** | 92.6% |
| **$RMSE_{ln(Kr)}$<10** | 96.0% | 97.3% | 100% | 100% | **100%** |

**Table S5. Pairwise comparisons of soil hydraulic parameters across texture groups.**

| Parameter | Comparison | Adjusted P | Cliff's δ | Source bootstrap 95% CI |
|---|---|---|---|---|
| $A$ | Sand–Loam | $1.92 \times 10^{-8}$ | 0.652 | [0.387, 0.886] |
| $A$ | Sand–Silts | $1.35 \times 10^{-16}$ | 0.784 | [0.632, 0.927] |
| $A$ | Sand–Clay | $1.01 \times 10^{-9}$ | 0.802 | [0.622, 0.959] |
| $\beta$ | Sand–Silts | $2.34 \times 10^{-8}$ | 0.547 | [0.268, 0.693] |
| $\beta$ | Sand–Clay | 0.00965 | 0.421 | [0.086, 0.764] |
| $\beta$ | Loam–Silts | 0.00151 | 0.406 | [−0.032, 0.702] |
| $\theta_r$ | Sand–Loam | $6.07 \times 10^{-9}$ | −0.596 | [−0.785, −0.352] |
| $\theta_r$ | Sand–Silts | $8.71 \times 10^{-5}$ | −0.442 | [−0.634, −0.102] |
| $\theta_r$ | Sand–Clay | $8.31 \times 10^{-13}$ | −0.880 | [−0.986, −0.664] |
| $\theta_r$ | Loam–Silts | 0.0282 | 0.372 | [0.034, 0.698] |
| $\theta_r$ | Loam–Clay | 0.0209 | −0.614 | [−0.833, −0.311] |
| $\theta_r$ | Silts–Clay | $4.11 \times 10^{-5}$ | −0.744 | [−0.918, −0.507] |

**Table S6. Sensitivity of relative hydraulic conductivity predictions to integration bounds.** The wet and dry head cutoffs are denoted by $H_w$ and $H_d$, respectively. RMSE is calculated for ln($K_r$). Percentages indicate the proportion below each RMSE threshold.

| $H_w$ | $H_d$ | Mean $RMSE_{ln(Kr)}$ | RMSE < 2.5 | RMSE < 5 | RMSE < 10 |
|---|---|---|---|---|---|
| 0.5 | 50,000 | 2.502220 | 61.74% | 91.28% | 100.00% |
| 0.5 | 100,000 | 2.500597 | 61.74% | 91.28% | 100.00% |
| 0.5 | 500,000 | 2.500233 | 61.74% | 91.28% | 100.00% |
| 1 | 50,000 | 2.431559 | 65.77% | 92.62% | 100.00% |
| 1 | 100,000 | 2.430053 | 65.77% | 92.62% | 100.00% |
| 1 | 500,000 | 2.429530 | 65.77% | 92.62% | 100.00% |
| 5 | 50,000 | 2.543340 | 60.40% | 92.62% | 100.00% |
| 5 | 100,000 | 2.542685 | 60.40% | 92.62% | 100.00% |
| 5 | 500,000 | 2.542630 | 60.40% | 92.62% | 100.00% |